\documentclass[12pt]{article}

\usepackage{hyperref}
\usepackage{amsmath, amssymb, graphics}
\usepackage{slashed}
\newcommand{\mathsym}[1]{{}}

\def\id{\protect{{1 \kern-.28em {\rm l}}}}
\usepackage{color}

\def\be{\begin{eqnarray}}
\def\ee{\end{eqnarray}}
\def\p{{\partial}}

\makeatletter
\renewcommand\section{\@startsection {section}{1}{\z@}%
{-3.5ex \@plus -1ex \@minus -.2ex}%
{2.3ex \@plus.2ex}%
{\normalfont\large\bfseries}}
\renewcommand\subsection{\@startsection{subsection}{2}{\z@}%
{-3.25ex\@plus -1ex \@minus -.2ex}%
{1.5ex \@plus .2ex}%
{\normalfont\normalsize\bfseries}}

\makeatother

\def\Tr{{\rm Tr}}
\def\STr{{\rm STr}}

\def \foot {\footnote}
\def \bi{\bibitem}

\def \ha {{1 \over 2}}
\def \td {\tilde}
\def \ci{\cite}

\def\z{\zeta}

\def\a{\alpha}
\def\b{\beta}
\def\e{\varepsilon}
\def\p{\phi}

\def \del{\partial}
\def \a {\alpha}

\def\g{\gamma}
\def\s{\sigma}
\def\z{\zeta}

\def\ov{\over}

\def\b{\beta}
\def\l{\lambda}

\def \k {\varkappa}
\def\foot{\footnote}

\def \ci {\cite}

\def \Tr {{\rm Tr}}
\def \P {\Phi}
\def \l {\lambda}

\def \td {\tilde}

\def \m {\mu}

\def\e{\epsilon}
\def \bi{\bibitem}
\def \la {\label}
\def \l {\lambda}
\def\foot{\footnote}

\def \sql {{\sqrt \l}}
\def \adss {$AdS_5 \times S^5~$ }
\newcommand{\rf}[1]{(\ref{#1})}
\def \ov {\over}

\def\cc{\circ}

\def \ha{{1\ov 2}}
\def \r {\rho}
\def \no {\nonumber}

\def \del {\partial}

\def \bi{\bibitem}
\def \la {\label}
\def \l {\lambda}
\def\foot{\footnote}

\def \sql {{\sqrt \l}}

\def \adss {$AdS_5 \times S^5$\ }

\def \p {\phi}
\def \r {\rho}

\def \ov {\over}

\def \varpi {{\rm w}}

\def \ep {\epsilon}

\def \DD {{\rm D}}

\def\Tr{{\rm Tr}}

\def \s {\sigma}

\def \n {\nu}
 \def \sql {\sqrt{\lambda}}
\def \vp {\varphi}

\def \ha {{{\textstyle{1 \ov2}}}}
\def \fo {{\textstyle{1 \ov4}}}

\def \sql {{\sqrt{\l}}}

\def \sql {\sqrt{\l}}

  \def \eqref {\rf}
\def \ads {$AdS_3 \times S^3$\ }
\def \adss {$AdS_5\times S^5$\ }
\def \adt {$AdS_2 \times S^2$\ }
\def \iffa {\iffalse}

\begin{document}

\overfullrule=0pt
\parskip=2pt
\parindent=12pt
\headheight=0in \headsep=0in \topmargin=0in \oddsidemargin=0in

\vspace{ -3cm}
\thispagestyle{empty}
\vspace{-1cm}

\rightline{Imperial-TP-AT-2014-02}
\rightline{HU-EP-14/10}

\begin{center}
\vspace{1cm}
{\Large\bf
On deformations of $AdS_n \times S^n$ supercosets
\vspace{0.2cm}
}
\vspace{1.5cm}

{B. Hoare$^{a,}$\footnote{ben.hoare@physik.hu-berlin.de}, R. Roiban$^{b,}$\footnote{radu@phys.psu.edu} \ and A.A. Tseytlin$^{c,}$\footnote{Also at Lebedev Institute, Moscow. tseytlin@imperial.ac.uk }}\\

\vskip 0.3cm

{\em $^{a}$ Institut f\"ur Physik, Humboldt-Universit\"at zu Berlin, \\ Newtonstrasse 15, D-12489 Berlin, Germany}
\vskip 0.08cm

{\em $^{b}$Department of Physics, The Pennsylvania State University,\\
University Park, PA 16802 , USA}
\vskip 0.08cm

{\em $^{c}$ The Blackett Laboratory, Imperial College, London SW7 2AZ, U.K.}

\vspace{.2cm}

\end{center}

{\baselineskip 11pt
\begin{abstract}
\noindent
We study the deformed $AdS_5 \times S^5$ supercoset model of arXiv:1309.5850
which depends on one parameter $\k$ and has classical quantum group symmetry.
We confirm the conjecture that in the ``maximal'' deformation limit, $\k\to \infty$, 
this model is T-dual to ``flipped'' double Wick rotation 
of the target space $AdS_5 \times S^5$, i.e. $dS_5 \times H^5$ space supported
by an imaginary 5-form flux. In the imaginary deformation limit, $\k \to i$,
the corresponding target space metric is of a pp-wave type and thus the
resulting light-cone gauge S-matrix becomes relativistically invariant.
Omitting non-unitary contributions of imaginary WZ terms, we find that this
tree-level S-matrix is equivalent to that of the generalized sine-Gordon model
representing the Pohlmeyer reduction of the undeformed $AdS_5 \times S^5$
superstring model. We also study in some detail similar deformations of the
$AdS_3 \times S^3$ and $AdS_2 \times S^2$ supercosets. The bosonic part of the
deformed $AdS_3 \times S^3$ model happens to be equivalent to the symmetric
case of the sum of the Fateev integrable deformation of the $SL(2)$ and $SU(2)$
principal chiral models, while in the $AdS_2 \times S^2$ case the role of the
Fateev model is played by the 2d ``sausage" model. The $\k=i$ limits are again
directly related to the Pohlmeyer reductions of the corresponding $AdS_n \times
S^n$ supercosets: (2,2) super sine-Gordon model and its complex sine-Gordon
analog.  We also discuss possible deformations of $AdS_3 \times S^3$ with more
than one parameter.
\end{abstract}
}
\newpage

\tableofcontents

\setcounter{footnote}{0}
\setcounter{section}{0}

\def \rq {{\rm q}}
\def \qq {{\rm q}}
\def \ll {{\ell }}
\def \rr {{\rm r}} \def \vvp {\chi}
\def\draftnote#1{{\color{red} #1}}
\def \rd {{\rm d}} \def \rc {{\rm c}}
\def \dd {{\rm d}} \def \cc {{\rm c}}
\def \na {\nabla}
\def \te {\textstyle}
\def \br {{\bar \r}}
\renewcommand{\theequation}{1.\arabic{equation}}
\setcounter{equation}{0}
\section{Introduction }

The integrability of the \adss superstring theory provides an important tool
for finding its spectrum~\ci{beir}.  Given an integrable sigma model one may
construct closely related integrable models by applying T-duality
transformations (see, e.g., \ci{lm,frt,fa,rtw,bei}).  Recently, a novel example
of a one-parameter integrable deformation of the \adss supercoset model, not
related to T-duality was found in \ci{dmv} (following earlier constructions of
\ci{dmv1,k1,k2}).  In \ci{abf} the coordinate form of the bosonic part of the
corresponding string action was worked out and the background string metric and
the NS-NS 2-form were explicitly determined.  This 10d background has the
$SO(2,4) \times SO(6)$ symmetry of \adss broken to its Cartan subgroup
$[U(1)]^6$ and thus its dual gauge theory interpretation is not immediately
clear.

The deformed string model \ci{dmv} is parametrized by the string tension
$T_0\equiv g= { \sql \ov 2 \pi} $ and a real deformation parameter $\eta \in
[0, 1)$.  It is useful to also introduce related parameters $\k\in [0,\infty)$
and $q$ as in \ci{abf}
\be
\k= { 2 \eta\ov 1- \eta^2} \ , \ \ \ \ \ \ \ \ \ \ q= e^{- \nu/g} \ , \ \ \ \
\ \ \ \ \ \nu= {2\eta\ov {1 + \eta^2}} = {\k \ov \sqrt{1 + \k^2} }\ . \la{1} \ee
An interesting feature of the model of \ci{dmv} is its classical $q$-deformed
symmetry, suggesting that it is more symmetric than is apparent from its
Lagrangian description.\foot{Similar models with classical $q$-deformed
symmetry were discussed in \ci{km,jor1}.} Remarkably, the corresponding
tree-level light-cone (bosonic) S-matrix matched \ci{abf} the S-matrix with
$q$-deformed centrally-extended $[psu(2|2)]^2$ symmetry \ci{bk,beis,hhmphase} 
with real deformation parameter~$q$.

This leaves many open questions.  In particular, it is not clear whether this
deformation should have an interesting target space interpretation or if it is
just a member of a universality class of models with effectively equivalent
classical integrable structure, but related to the original undeformed one by
non-local transformations making the resulting quantum theories inequivalent.
Another interesting question is about the existence and properties of a gauge
theory dual to string theory in the deformed geometry. 

Our aim here will be to explore this deformed model by studying its simple
limits and low-dimensional analogs.  In particular, we shall consider in detail
the following two formal limits:

(i) $ \eta=1 $ or $\k=\infty$ ($q=e^{-1/g}$) and (ii) $ \eta=i $ or $\k=i$
($q=e^{i \infty/g}$).  \\ It turns out that in the first ``maximal
deformation'' limit the deformed 10d metric becomes closely related (T-dual) to
a flipped ``double Wick rotation"\footnote{This double Wick rotation of the target space 
should not be confused with the double Wick rotation on the world sheet used
to construct the mirror model of the light-cone gauge-fixed string theory.}
of the \adss space -- $dS_5 \times H^5$,
where $dS_5$ is the de Sitter space (whose euclidean continuation is $S^5$) and
$H^5$ is the 5d hyperboloid (which is the euclidean continuation of $AdS_5$).
This proves the conjecture of \ci{dmv} that the deformation effectively
interpolates between \adss and $dS_5 \times H^5$ spaces.  $dS_5 \times H^5$ is
a formal solution of type IIB supergravity supported by an imaginary self-dual
5-form flux \ci{hul} implying that the corresponding world-sheet theory is
non-unitary.

In the second ``imaginary deformation'' limit (combined with a particular
rescaling of coordinates) the 10d metric becomes that of a pp-wave background
with a curved transverse part.  The corresponding light-cone gauge string
action takes a form reminiscent of the action of the Pohlmeyer reduced (PR)
theory for the \adss superstring\foot{The Pohlmeyer reduction is based on
solving the string Virasoro conditions in terms of new ``current'' variables
related through derivatives (i.e. non-locally) to the original string
coordinates and then writing down the action for the new unconstrained
variables.} \ci{gt1,ms,gt2} but with additional imaginary $B$-field (WZ) terms,
implying that unitarity is broken.  The resulting light-cone gauge S-matrix is
then {\it relativistically invariant} and, ignoring the imaginary WZ term
contribution, happens to be the same as the tree-level PR S-matrix found in
\ci{ht1,ht2}.

We shall also study the direct 6d and 4d analogs of the deformed \adss model,
which may be interpreted as deformations of the \ads \ci{bab} and \adt \ci{beb}
supercosets.  The corresponding metrics are direct sums of deformed $AdS_n$ and
deformed $S^n$ metrics and are simply given by truncations of the corresponding
parts of the deformed 10d metric of \ci{abf}.  The integrability of the
resulting 6d and 4d classical string models is inherited from the integrability
of 10d model.

As we shall explain below, the corresponding bosonic integrable models were
identified before in a different guise: the one-parameter $\k$-deformation of
the $S^3$ metric (corresponding to a deformation of the $SO(4)/SO(3)$ coset
following \ci{dmv1}) is a special ``left-right symmetric'' case of the Fateev
2-parameter deformation of the $SU(2)$ principal chiral model \ci{fat}, the
classical integrability of which was proved in \ci{luk}.  Similarly, the
$\k$-deformation of the $S^2$ metric (found also as the $SU(2)/U(1)$ coset
deformation in \ci{dmv1}) is the same as the 2d ``sausage'' model of \ci{foz},
for which the classical Lax pair was given in \ci{luk}.

We shall show that the general 2-parameter Fateev model \ci{fat} is the same as
the $SU(2)$ case of the 2-parameter family of classically integrable
``bi-Yang-Baxter'' sigma models constructed in \ci{k2,k3}.\foot{The
``diagonal'' ($\a=\b$) limit of the 2-parameter $SU(2)$ YB model of \ci{k3} is
the same as the $SO(4)/SO(3)$ coset deformation of \ci{dmv1} or the
``symmetric'' one-parameter case of the Fateev model.  The one-parameter
($\b=0$) case is the same as the original Yang-Baxter model of \ci{k1,k2}
which, in the $SU(2)$ case, is the squashed $S^3$ model known to be integrable
since \ci{che}.} This suggests the existence of a two-parameter deformation of
the \ads supercoset model with the bosonic part being given by the sum of the
$SU(2)$ Fateev model and its $SL(2,R)$ analog.  Furthermore, the Fateev model
admits an integrable extension \ci{luk} to the presence of a WZ term or
non-zero $B$-field coupling, implying that it might be possible to construct a
3-parameter deformation of the \ads supercoset \ci{kaz} with non-zero NS-NS
$B$-field coupling (containing as a special case the $SL(2,R) \times SU(2) $
WZW model).

For these low-dimensional \ads and \adt models one may also study the special
$\k=\infty$ and $\k=i$ limits.  In particular, in the $\k=i$ limit the
resulting pp-wave model turns out to be closely related to the one in
\ci{mm,rt,bs}.  After completing these pp-wave 4d and 6d metrics to
supergravity solutions we will find that in light-cone gauge they reduce, in
the 4d case, to the $(2,2)$ supersymmetric sine-Gordon model which is
equivalent to the PR model for the \adt superstring \ci{gt1}, and, in the 6d
case, to a fermionic extension of the sum of the complex sine-Gordon and
complex sinh-Gordon models which is equivalent to the PR model \ci{gt2} for the
\ads superstring and has hidden (4,4) supersymmetry.  In the \adt case, we will
also construct explicitly the quadratic fermion terms in the \adt analog of the
deformed supercoset action of \ci{dmv} for $\k=i$ and show that it also
reproduces the PR model for the \adt superstring \ci{gt1}. A similar analysis
should be possible for the \ads and \adss cases as well.

We shall start in section 2 with a review of the 10d $\k$-dependent metric and
$B$-field background corresponding \ci{abf} to the deformed \adss model of
\ci{dmv} and then consider the special limits of $\k=\infty$ and $\k=i$ and
low-dimensional truncations.

The deformed \ads case will be discussed in detail in section 3, where we
explain the equivalence of the $\k$-deformed $S^3$ metric to the symmetric case
of the Fateev model and discuss the relation between the $\k=i$ limit of the
deformed \ads background and the Pohlmeyer reduced model for the original
undeformed \ads superstring theory.

Section 4 is devoted to the \adt case. We shall start with the deformed \adt
supercoset Lagrangian constructed following \ci{dmv} and show that its bosonic
part corresponds to the 4d truncation of the 10d $\k$-deformed metric.  We
shall then consider the $\k=i$ case and show its equivalence in the light-cone
gauge to the PR model for the undeformed \adt supercoset.
We shall also demonstrate that the deformed sigma model is one-loop UV finite
when expanded near a BMN-type geodesic. 

In Appendix A we demonstrate the equivalence between the Fateev model \ci{fat}
and the 2-parameter $SU(2)$ bi-Yang-Baxter sigma model of \ci{k2,k3}. Appendix
B presents a review of the 4-parameter Lukyanov sigma model \ci{luk}, which
generalizes the Fateev model introducing a $B$-field coupling.  Appendix C
contains the details of the construction of the deformed supercoset action for
\adt at $\k=i$ and the demonstration of its equivalence to the Pohlmeyer
reduced model for the undeformed \adt superstring.

\renewcommand{\theequation}{2.\arabic{equation}}
\setcounter{equation}{0}
\section{Deformed \adss model and its limits }

The deformed \adss string action may be written as \ci{dmv,abf}
\be \la{2}
S=\ha \hat T \int d^2 \s \ \big( L_G + L_B + L_{\rm ferm} \big) \ , \ \ \ \ \ \ \ \
\hat T = g ( 1 + \k^2)^{1/2}
\ ,
\ee
where $\hat T$ is the effective string tension\foot{We use the definition of
the tension from \ci{abf}; the choice in \ci{dmv} was $ \hat T = g ( 1 +
\k^2)^{3/2}$.} $L_G$ is the string metric $G$ part and $L_B$ is the 2-form $B$
or WZ part. The fermionic terms should contain couplings to $e^\P F_k$
($k=1,3,5$) where $\P$ is the dilaton (which is non-constant for generic $\k$)
and $F_k$ are RR fluxes. $\P$ and $F_k$ (which are presently unknown) should
supplement $G$ and $ B$ to give a type IIB supergravity solution to ensure
conformal invariance of the model as suggested by the fermionic kappa-symmetry
of the deformed action \ci{dmv}.

Explicitly, the deformed analog of the $AdS_5$ metric is \ci{abf}
\be
&& ds^2_{A_5} = - h(\r) dt^2 + f(\r) d\r^2 + \r^2 \, \big[\, v(\r, \z)\, ( d\z^2 + \cos^2 \z\, d\psi_1^2) + \sin^2 \z\, d\psi_2^2\,\big] \ , \la{3} \\
&&h= { 1 +\r^2 \ov 1 - \k^2 \r^2} \ , \ \ \ \ f= { 1 \ov (1 +\r^2)( 1 - \k^2 \r^2)} \ , \ \ \ \
v= { 1 \ov 1 + \k^2 \r^4 \sin^2 \z } \ .\la{4} \ee
For $\k=0$ this is the standard global $AdS_5$ metric with embedding coordinates
$X_0 + i X_5 = \sqrt{ 1 + \r^2} e^{it}, \ X_1 + i X_2 = \r\cos\z\, e^{i\psi_1 }, \ X_3 + i X_4 = \r\sin\z\, e^{i\psi_2}$.
The deformed $S^5$ metric is found by a simple analytic continuation $\r\to i r$ and reversing the overall sign of the metric:
\be
&& ds^2_{S_5} = \td h(r) d\vp^2 + \td f(r) dr^2 + r^2\, \big[\, \td v(r, \theta)\,
( d\theta^2 + \cos^2 \theta\, d\p_1^2) + \sin^2 \theta \, d\p_2^2\,\big] \ , \la{5} \\
&&\td h= { 1 -r^2 \ov 1 + \k^2 r^2} \ , \ \ \ \ \td f= { 1 \ov (1 - r^2)( 1 + \k^2 r^2)} \ , \ \ \ \
\td v= { 1 \ov 1 + \k^2 r^4 \sin^2 \theta } \ .\la{6} \ee
The non-zero $B_{mn}$ components in the two subspaces are
\be\la{7}
B_{\psi_1 \z} = \ha { \k\, \r^4 \sin2 \z\ } v (\r,\z) \ ,\ \ \ \ \ \ \ \ \ \ \ \ B_{\p_1 \theta} =- \ha { \k \, r^4 \sin 2\theta\ } \td v (r,\theta)
\ . \ee
The deformed metrics \rf{3}, \rf{5} have only the $[U(1)]^3$ Cartan subgroups of the original $SO(2,4)$ and $SO(6)$
as their surviving symmetry. 
The deformed target space background should not have (for generic value of $\k$) 
any manifest supersymmetry but the string model \rf{2} of \ci{dmv} should have hidden symmetries due to its integrability. 

Assuming that the above 10d metric and $B$-field background can indeed be
completed to a full type II supergravity solution\foot{\label{footnoteGS} This is a non-trivial assumption: it does not seem likely that
a generic NS-NS ($G,B)$ background may be completed by a dilaton and RR fluxes
to a type II supergravity solution. In the present case this is expected provided
the deformed supercoset model of \ci{dmv} does indeed
have an interpretation as a GS action in a type IIB supergravity background \ci{gh}. This is supported, in particular,
by its kappa-symmetry \ci{dmv} and the special limits of $\k=\infty$ and $\k=i$ when this is the case as discussed below.}
the corresponding dilaton should satisfy the following equation
\be \la{8}
R + 4 \na^2 \P - 4 (\na_m\P)^2 - \te {1\ov 12} H^2_{mnk} =0 , \ \ \ {\rm i.e.} \ \ \ \
\na^2 e^{-\P} + \fo ( R - \te {1\ov 12} H^2_{mnk}) e^{-\P}=0 \ . \ee
This equation does not appear to have a simple rational solution for the above background,
suggesting that the RR fluxes $F_k$ should also have a complicated form, such that $e^{\P} F_k$ is rational.
This is required to match the structure of the fermionic terms in \rf{2} that should
have a rational dependence on coordinates,
as implied by the construction of \ci{dmv} for the coordinate parametrization used in \ci{abf}.

The deformed string metric \rf{3} has a curvature 
singularity at $\rho_*=1/\k$; for larger values the radial coordinate $\rho$ becomes time-like, 
suggesting that strings are confined to the region $0\le \rho \le 1/\k$. Thus for $\k\not=0$ 
the metric \eqref{3} no longer has a boundary (which reappears at $\r=\infty$ if $\k=0$).
It is an open question if/how string theory resolves this singularity.
It would also be interesting to understand in detail whether string theory in the deformed geometry \eqref{4} supplemented by the 
required fluxes and dilaton has a gauge theory dual. In the absence of non-abelian isometries it should not have conformal or even 
Lorentz symmetry (and of course no manifest supersymmetry).\foot{A singularity
in the Einstein frame metric would suggest that the UV limit of this theory is in some sense not well-defined
perhaps due to deformation by an irrelevant operator. A similar conclusion may be reached by analysing the deformation in 
coordinates in which the metric reduces to that of the \adss in the Poincar\'e patch as $\k\rightarrow 0$.}

\subsection{$\k=\infty$ limit }

Let us now consider the ``maximal deformation'' limit, $\k\to \infty$ (or $\eta=1$), in the string action~\rf{2}.
If we formally take this limit in the metric \rf{3}, \rf{4} we get
\be \la{9}
&& ds^2_{A_5, \k\to \infty} = {1 \ov \k^2} \bar {ds}^2_{A_5} \ , \ \ \ \ \ \
\bar {ds}^2_{A_5}=
(1+ \br^2) dt^2 - (1+ \br^2)^{-1} d\br^2 + \br^2 ds^2_3 \ , \\
&& ds^2_3 = { d\z^2 \ov \sin^2 \z} + \cot^2 \z\, d\psi_1^2 + \br^{-4} \sin^2 \z\, d \bar \psi_2^2 \ , \la{10} \\
&& \br \equiv \r^{-1} \ , \ \ \ \ \ \ \bar \psi_2\equiv \k \psi_2 \ . \la{11} \ee
Here we have redefined $\r$ and $\psi_2$ (so that $\bar \psi_2$ is non-compact for $\k \to \infty$).
The corresponding $B$ term in \rf{2}, \rf{7} becomes a total derivative in this limit and can be ignored.
Performing a formal T-duality along the $ \bar \psi_2$ direction
and changing the coordinate $\z\to y$ we find that the
T-dual metric
becomes ($ y\equiv \ln \tan { \z\ov 2} $)\foot{We assume that the dual coordinate $\td {\bar \psi}$ is rescaled by $\hat T \k^{-2} $,
where $\hat T$ is the string tension in \rf{2}.}
\be
{\td {\bar{ds}}}{}^2_{A_5} = ds^2_{dS_5} = - (1+ \br^2)^{-1} d\br^2 +
(1+ \br^2) dt^2 + \br^2 \big ( { dy^2 } + \sinh^2 y \, d\psi_1^2 + \cosh^2 y \, d \td {\bar \psi}_2^2\big) \ .
\la{12} \ee
This metric is the same as the metric of the {\it de Sitter} ($dS_5$) space
with $\br$ now playing the role of the
time coordinate.\foot{Setting
$\br = \sinh {\xi} $ we get
$ ds^2_{dS_5} = - d\xi ^2 +
\cosh^2 \xi \, dt^2 + \sinh^2 \xi\, \big ( { dy^2 } + \sinh^2 y \, d\psi_1^2 + \cosh^2 y \, d \td {\bar \psi}_2^2\big) $.
The scalar curvature of this metric is $R=20$. One can introduce global coordinates in $R^{1,5}$ such that
this metric becomes that of the positive-curvature surface $-X_0^2 + X_1^2 + X_2^2 + X_3^2 + X_4^2 + X_5^2 =1$.}
Thus, while at $\k=0$ the metric \rf{3} is that of the
negative-curvature $AdS_5$ space, at $\k=\infty$ it is T-dual
to the positive-curvature $dS_5$ metric.

\def \ro {{\varrho}} 
\def \tem {{\rm t}} 

Similarly, the $\k \to \infty$ limit of the
deformed $S^5$ metric \rf{5} becomes (after the analogous coordinate transformations, $\bar r = r^{-1}$, etc.,
and T-duality)
the metric of the negative-curvature euclidean $AdS_5$ space or the
hyperboloid $H^5$ ($ x\equiv \ln \tan { \theta\ov 2} $):\foot{This metric of $H^5$ is written in the hyperbolic slicing; its
scalar curvature is $R=-20$. Indeed, one can introduce global coordinates in $R^{1,5}$ such that
this metric becomes the metric of the negative-curvature surface $-X_5^2 + X_0^2 + X_1^2 + X_2^2 + X_3^2 + X_4^2 =-a^2$.
Similarly, the de Sitter space corresponds to the 
positive-curvature surface $-X_0^2 + X_5^2 + X_1^2 + X_2^2 + X_3^2 + X_4^2 =a^2$ (we set the radius $a=1$ in above expressions). 
One may choose the static coordinates 
as $X_0= \sqrt{a^2 - \ro^2}\sinh ({ \tem/a}) , \ X_5 = \sqrt{a^2 - \ro^2}\cosh ( { \tem /a}) , \ 
X_k = \ro n_k $ (where $n_k$ are the 3-sphere coordinates, $n_k n_k =1$) 
in which the $dS_5$ metric becomes 
$ds^2_{dS_5} = - ( 1- {\ro^2 /a^2}) d \tem^2 + ( 1- {\ro^2 /a^2})^{-1} d \ro^2 + \ro^2 d \Omega_3 ^2$. 
The analogous metric for $H^5$ is 
$ds^2_{H^5} = ( 1+ {\xi^2 /a^2}) d \p^2 + ( 1+ {\xi^2 /a^2})^{-1} d \xi^2 + \xi^2 d \Omega_3 ^2$. 
\label{stat}
}
\be
{\td {\bar {ds}}}^2_{S_5}= ds^2_{H^5} = (\bar r^2-1)^{-1} d\bar r^2 + (\bar r^2-1) d\vp ^2
+ \bar r^2 \big ( { dx^2 } + \sinh^2 x \, d\p_1^2 + \cosh^2 x \, d \td {\bar \p}_2^2\big) \ .
\la{121} \ee
We conclude that the $\k$-deformation
interpolates between $AdS_5 \times S^5$ at $\k=0$ and the T-dual of $dS_5 \times H^5$ at $\k=\infty$.
This effectively confirms the conjecture made in \ci{dmv}\foot{T-duality was not mentioned in \ci{dmv}
but at the level of the first-order formalism used in \ci{dmv1} it may be viewed as a kind of canonical transformation
(which is non-local in terms of the original coordinate fields).}
which was motivated by a similar interpolation between the $SU(2)/U(1)$ and $SU(1,1)/U(1)$ cosets observed in \ci{dmv1}.

It is interesting to note that $dS_5 \times H^5$ is also a double Wick rotation of $AdS_5 \times S^5$
combined with a $Z_2$ interchange of the factors (the euclidean rotation of $AdS_5$ is $H^5$ and the Minkowski version of $S^5$ is $dS_5$).
For that reason this space
also solves the (complexified) type IIB supergravity equations if it is
supported by an {\it imaginary} self-dual
5-form flux (the 5+5 Ricci tensor blocks should change
signs as compared to the \adss case, which is supported by a
real self-dual 5-form flux).\foot{Such a
solution
of a non-unitary analytic continuation of type IIB supergravity was discussed
earlier in \ci{hul}.}

Reversing the T-duality transformations along $ \psi_2$ and $\p_2$
we get a type IIB supergravity solution (with the metric in \rf{9}, \rf{10} and its $S^5$ counterpart) supported by an imaginary constant self-dual
$F_5$ flux and the following dilaton field (originating from the T-duality transformations) $ \P = \P_A + \P_S $
\be
( \P_A)_{\k=\infty} = - \ln( \bar \r^2 \cosh y) = \ln( \r^2 \sin \z) \ , \ \ \
(\P_S)_{\k=\infty} = - \ln( \bar r^2 \cosh x) = \ln( r^2 \sin \theta)\ . \ \ \ \la{13}
\ee
The fact that the $\k=\infty$ limit of the deformed background is a formal
solution of type IIB supergravity (with an imaginary $F_5$ flux)
verifies that the corresponding
limit of the deformed superstring action \ci{dmv} should be describing a 2d conformal theory.
However, the presence of an imaginary $e^\P F_5$ coupling in the fermionic part of the string action
suggests that the resulting world-sheet theory is likely to be {\it non-unitary}.

This non-unitarity is probably related to a special nature of the
limit $\k=\infty$ (or $\eta=1$): in
this case the quantum deformation parameter $q$ in \rf{1} approaches unity
in the perturbative string limit, $g \to \infty$.\foot{Since for $\k \to \infty$ the metric in \rf{9} scales as $\k^{-2}$
and the tension in \rf{2} goes as $\hat T \sim g \k $
to get a finite string action in this limit one would need to rescale $g$ by $\k$.
This will make $q$ in \rf{1} go to 1.
Alternatively, we may keep $g$ fixed and rescale the string coordinates to cancel the overall factor of $\k^{-1}$.
In this case
(in full analogy with the \adss case \ci{hu,bmn}, using the static coordinates of footnote 9)
we will end up with a pp-wave limit of the $dS_5 \times H^5$ background,
$
ds^2 = 4dx^+dx^- + ( x_r^2 + y_m^2) dx^+{}^2 + dx_r dx_r + dy_m dy_m $.
Then the 4+4 massive bosonic fluctuations found in light-cone gauge will be tachyonic.
Similarly, the fermionic mass terms (which will be imaginary due to the imaginary $F_5$ flux) 
will also correspond to non-unitary tachyonic modes. 
\label{foottach}
}

Interestingly, such a limit taken in the interpolating
S-matrix \ci{bk,hhmphase} with real $q$ formally
corresponds to the mirror theory S-matrix (cf. \ci{van})\foot{The mirror theory is found via a double Wick rotation on the world sheet 
in the light-cone gauge \ci{AF}. The mirror TBA was discussed for complex $q$ equal to root of unity
in \ci{alt}.} and in this context it is not clear why non-unitarity should appear for real $q$.
One possible resolution of this puzzle is to consider the light-cone
gauge-fixed string in the $dS_5 \times H^5$ background in static coordinates 
and formally interchange the world-sheet coordinates or, equivalently, the world-sheet energy and momentum.
The dispersion relation of the tachyonic modes discussed in footnote \ref{foottach} then becomes
the usual massive one. Assuming this prescription
also extends to the interaction
terms, the light-cone gauge-fixed mirror theory should then be equivalent to
the light-cone gauge-fixing of the string in the $dS_5 \times H^5$ background with the world-sheet
coordinates (and the corresponding conserved charges) formally interchanged.

There is also a more general perspective on this (non)unitarity issue. 
The deformed $AdS_5$ metric \rf{3} contains factors of $1-\k^2 \r^2$ 
implying that $\r$ is formally restricted to the interval $ 0 \leq \r < \k^{-1}$
($ \r = \k^{-1}$ is the curvature singularity). Continuing $\r$ beyond $\k^{-1}$ 
implies that $\r$ becomes time-like, while $t$ becomes space-like. Also, the RR fluxes may 
contain factors of $\sqrt{ 1-\k^2 \r^2}$ and so they may become imaginary for $ \r > \k^{-1}$.
This may be an indication that the ``unphysical" region $ \k^{-1} \leq \r < \infty $ is describing 
a non-unitary world-sheet theory. The $\k=\infty$ limit discussed above corresponds
to the case when the ``physical" region $ 0 \leq \r < \k^{-1}$ shrinks to a point while 
the ``unphysical" one extends to the whole half-line. 
Starting with the unitary light-cone gauge 
S-matrix found as in \ci{abf} in the ``physical" $ 0 \leq \r < \k^{-1}$ region and 
taking the formal limit $\k \to \infty$ corresponds effectively to switching to 
the S-matrix in the ``unphysical" region. It should be noted that the light-cone gauge fixing and $\k\to \infty$ 
limit need not commute. Indeed, the original BMN geodesic in the $ 0 \leq \r < \k^{-1}$ region 
may become complex in the ``unphysical" region, while the 
non-unitarity of the $dS_5 \times H^5$ S-matrix refers to the expansion near a different vacuum -- the real BMN type geodesic 
in static coordinates. This may be a resolution of the tension with unitarity of the mirror S-matrix. 

Finally, let us note also that this intriguing relation of \adss to its double Wick rotation $dS_5 \times H^5$
via the $\eta$-deformation
is potentially hinting at a more universal description in terms of complexification of the underlying
(super)group or in terms of its ``double'' (cf. \ci{k1}).

\subsection{$\k=i$ limit }

Even though the model of \ci{dmv} is defined for real $\eta$,
let us study the formal limit of $\eta \to i$ or $\k \to i$
as
it has some interesting features and, in particular, establishes a connection to the Pohlmeyer reduced
model for the original ($\k=0$) \adss superstring.

Directly setting $\k=i$ in the metrics \rf{3} and \rf{5} we
observe that $t$ and $\vp$ directions decouple
\be
ds^2_{A_5, \k= i } = - dt^2 + ds^2_{A\perp}
\ , \ \ \ \ ds^2_{A\perp} = { d\r^2\ov (1 +\r^2)^{2}} + \r^2 \Big[
{ d\z^2 + \cos^2 \z\, d\psi_1^2\ov 1 - \r^4 \sin^2 \z} + \sin^2 \z\, d\psi_2^2\Big]
\ , \la{44} \\
ds^2_{S_5, \k= i } = d\vp^2 + ds^2_{S\perp} \ , \ \ \ \ \ \
ds^2_{S\perp} =
{dr^2\ov (1-r^2)^2} + r^2 \Big[
{ d\theta^2 + \cos^2 \theta\, d\p_1^2\ov 1 - r^4 \sin^2 \theta} + \sin^2 \theta\, d\p_2^2\Big]
\ . \ \, \la{45}
\ee
Thus the 10d metric
factorizes as $R_t \times S^1_\vp \times M_A^4 \times M_S^4$.\,\foot{The scalar curvature of $M_A^4$
is $R=
\frac2{\r^2} \Big[2 \left(1-\r^4 \sin ^2\z\right)
-4 \left(\r^2+1\right)
-\frac{28 \left(\r^2+1\right)^2}{\left(1-\r^4 \sin ^2\z\right)^2}
+\frac{6 \left(3 \r^2+5\right) \left(\r^2+1\right)}{1-\r^4 \sin ^2\z}\Big]$.}
The $B$-field in \rf{7} becomes imaginary, implying that the resulting
string action will represent an integrable but non-unitary theory.
This is not surprising since for $\k=i$ the $q$-parameter in \rf{1} is complex
and thus the corresponding light-cone S-matrix becomes non-unitary (cf. \ci{bk,ht2,hmh,alt,hhmunit,abf}). 

Another indication that this limit is special is that the resulting light-cone S-matrix is {\it relativistically invariant}:
the decoupling of the two directions $t,\vp$
implies that the light-cone gauge fixing is straightforward
(as in flat and pp-wave space examples)
and does not break
2d Lorentz invariance.
As we shall see, this limit is closely related to the Pohlmeyer reduced
model \ci{gt1,gt2} for the undeformed \adss superstring which has a relativistic massive S-matrix.

As in the $\k\to \infty$ case discussed above,
there is, however, a subtlety to be addressed: in the formal substitution of $\k=i$ into the metric
we ignored the fact that the effective tension in \rf{2} vanishes.
To get a non-zero string action we thus
need to either (i) rescale $g$ (taking it to infinity as $\k\to i$ so that $\hat T$ in \rf{2} stays finite)
or (ii) keep $g$ fixed and compensate $\hat T$ going to zero by rescaling
string coordinates (as in the standard pp-wave limit, see, e.g., \ci{sft}).
Let us follow the first route but
also correlate $\k \to i$ with a rescaling of just $t$ and $\vp$ in \rf{3}, \rf{5}.
This allows us to define the $\k\to i$ limit in a non-trivial way,
so that the $t,\vp$ directions do not automatically decouple:
\be
\k^2 = -1 + s\, \epsilon^2 \ ,
\qquad
t = {\epsilon}^{-1} x^+ - \epsilon x^- \ ,
\qquad
\vp = {\epsilon}^{-1} x^+ + \epsilon x^-
\ , \ \ \ \ \ \ep\to 0 \ ,
\label{lim}
\ee
where $s$ is an arbitrary constant.
Then from \rf{3}, \rf{5}, \rf{7} we get the following pp-wave type 10d metric and $B$-field
\be
&&ds^2=
4 dx^+dx^- - s \big[ V(\a) +\td V(\b) \big] dx^+{}^2 + ds^2_{A\perp} + ds^2_{S\perp} \la{k} \ ,\\
&&
V(\a) = \sin^2 \a \ , \ \ \ \ \ \td V(\b) = \sinh^2 \b \ , \ \ \ \ \ \
\r\equiv \tan \a \ , \ \ \ \ r\equiv \tanh \b \ , \la{kk} \\
&&
ds^2_{A\perp} = { d\a^2} + \tan^2\a\, \big[
{ d\z^2 + \cos^2 \z\, d\psi_1^2\ov 1 - \tan^4 \a\, \sin^2 \z} + \sin^2 \z\, d\psi_2^2\big]
\ , \la{k4} \\ &&
ds^2_{S\perp} =
{d\b^2} + \tanh^2\b\, \big[
{ d\theta^2 + \cos^2 \theta\, d\p_1^2\ov 1 - \tanh^4 \b\, \sin^2 \theta} + \sin^2 \theta\, d\p_2^2\big]
\ ,\ \ \ \ \la{k5} \\
&& B_{\psi_1\zeta} = i\, \frac{\tan^4\a\, \sin \zeta\, \cos \z }{1-\tan^4\a\, \sin^2\zeta}
\ , \ \ \ \ \ \ \ \ \ \ \ \ \
B_{\phi_1\theta} = - i\, \frac{\tanh^4\b\, \sin \theta\, \cos \theta }{1- \tanh^4 \b \, \sin^2\theta} \ , \qquad \la{bi}
\ee
where the 4d ``transverse'' metrics \rf{k4}, \rf{k5} are the same as in \rf{44}, \rf{45} after the coordinate transformations.
Fixing the light-cone gauge $x^+ = \mu \tau$ (and ignoring the fermions) then gives the direct
sum of two bosonic relativistic interacting
integrable massive models.\foot{Expanding in small $\a$ and $\b$ gives a theory of 4+4 massive fields.}
Note that $s=0$ is the case of the naive $\k=i$ limit in \rf{44}, \rf{45} where the light-cone gauge theory had no potential.
The case of $s <0$ leads to negative-definite potential, so in what follows we
set $s=1$.\foot{The norm of $s$ does not matter as it can be absorbed into a rescaling of $x^+$ and $x^-$ or $\mu$.}

The resulting theory of $4+4$ massive bosons looks very similar
to the bosonic truncation of the generalized sine-Gordon model that
appeared as the Pohlmeyer reduction (PR) of the \adss superstring \ci{gt1}.
One may wonder if there is a gauge fixing of the gWZW model representing
the PR of the \adss string \ci{gt1} that leads directly to this light-cone theory.
This may seem unlikely for several reasons:
(i) the PR action is real, while the $\k=i$ limit leads to an imaginary WZ term;
(ii) the metric of the $G/H$ gWZW model with non-abelian $H$, found after solving for the $H$-gauge field,
should have no isometries \ci{gt1}, while here we have
four $U(1)$ isometries;
(iii) there is no $B$-field coupling in the gWZW model with maximal subgroup $H$ gauged \ci{gt1},
while the $\k=i$ limit leads to a non-zero (and imaginary) WZ term;
(iv) the metric of the $G/H$ gWZW model with non-abelian $H$ does not admit a perturbation theory around a simple
vacuum, while here there is a well-defined expansion around the $\a=\b=0$ point.

There are, however, hints of a close connection between the two theories. As observed earlier,
to retain a finite string tension while taking $\k \to i$
we also need to take $g \to \infty$. Therefore,
it is natural to expect that the S-matrix for the massive
excitations of the light-cone theory originating from \rf{k}
should be related to the strong coupling
limit of the interpolating S-matrix \ci{bk,hhmphase} with $q$ as a phase.\footnote{There are different ways to take the strong coupling limit of the interpolating S-matrix \ci{bk,hhmphase} (see, e.g. \ci{beis}). In particular, in \ci{alt} it was pointed out that, depending on the scaling of the world-sheet momentum, one can arrive at either a massless dispersion relation, which should correspond to $s=0$ in \eqref{lim}, or the massive PR dispersion relation, corresponding to $s=1$.}
This S-matrix is not unitary, which
is a reflection of
the non-reality of the light-cone gauge Lagrangian mentioned above.

The connection between the interpolating S-matrix in this limit
and the Pohlmeyer reduction of the \adss superstring was observed in \ci{beis} and discussed in detail
in \ci{ht2,hmh}. While there is no precise agreement, at tree-level the S-matrix of the PR theory is given by the
parity-even (unitary) piece of the interpolating S-matrix.\foot{Note that an alternative gauge fixing,
bypassing the use of the metric of the $G/H$ gWZW model, was used in \ci{ht1} to set up a perturbative expansion
around the trivial vacuum.}
This can also be seen explicitly in the expansion to quartic order
of the light-cone gauge theory corresponding to \rf{k}--\rf{bi}:
\be
L = & \no -{\del_i\a\del^i\a} - (\a^2+{2\a^4 \ov 3}) \Big(
\del_i \z\del^i \z + \cos^2 \z\, \del_i\psi_1\del^i\psi_1 + \sin^2 \z\, \del_i\psi_2\del^i\psi_2\Big) -\m^2( \a^2-{\a^4 \ov 3})
\\ & \no - {\del_i\b\del^i\b} -(\b^2-{2\b^4\ov 3}) \Big(
\del_i\theta\del^i\theta + \cos^2 \theta\, \del_i\p_1\del^i\p_1 + \sin^2 \theta\, \del_i\p_2\del^i\p_2\Big)
-\m^2( \b^2 + {\b^4 \ov 3} )
\\ & +\ i \a^4\sin2\z \, \e^{ij}\del_i\psi_1 \del_j \z - i \b^4 \sin2\theta\, \e^{ij}\del_i \phi_1 \del_j \theta + O(\a^6) + O(\b^6)\ .\la{pr}
\ee
Introducing the fields \def\ypr{Z}\def\zpr{Y}
\be
&& \ypr_1 + i \ypr_2 =\a \cos\z\, e^{i\psi_1} \ , \qquad \ypr_3 + i \ypr_4 =\a \sin \z\, e^{i \psi_2} \ ,
\\ &&
\zpr_1 + i \zpr_2 =\b \cos\theta\, e^{i\p_1} \ , \qquad \ \zpr_3 + i \zpr_4 =\b \sin \theta\, e^{i \phi_2} \ ,
\ee
we find that the metric part of \rf{pr} (the first two lines) matches the quartic bosonic
terms of the PR action in \ci{ht1},
and thus the corresponding tree-level S-matrix (with non-unitary $B$-field terms omitted)
should match the tree-level PR theory S-matrix in \ci{ht1}.
At the same time, while the tree-level PR S-matrix of \ci{ht1} did not satisfy the standard classical Yang-Baxter equation,
the S-matrix corresponding to the above light-cone gauge theory with the imaginary $B$-field terms included will satisfy it
(in agreement with the classical integrability \ci{dmv} of the deformed theory for any $\k$).

It is also interesting to note that deforming the $AdS_5$ metric and taking $\k \to i$
gives rise to a model that is similar to
the PR of the string on $R \times S^5$, and vice versa -- the $\k=i$ limit of the deformed $S^5$ is similar to the PR of the
string on $AdS_5 \times S^1$.
Thus the roles of $AdS_5$ and $S^5$ appear to be interchanged
as we move from $\k=0$ to $\k=i$. Interestingly, this is a feature that was observed in
the S-matrix picture via an analysis of the scattering of bound states
in the interpolating theory with $q$ being a phase \ci{hhmbs}.

In \ci{hhmunit} it was claimed that the S-matrix for the physical states of the PR model should be
given by the
vertex-to-IRF transformation of the interpolating S-matrix with $q$ being a
phase. The resulting S-matrix is unitary and also has the perturbative tree-level S-matrix of \cite{ht1}
as a limit.
It remains to be seen if this transformation can be lifted precisely to a relation between the ``pp-wave'' model \rf{k}--\rf{bi} and the PR model of \ci{gt1}.

As we shall demonstrate below in sections 3 and 4, upon dimensional
reduction to deformations of the \ads and \adt theories the relation between the $\k=i$ deformed string theory
and the PR model for the undeformed theory
becomes much more straightforward.

\subsection{Consistent truncations to low-dimensional models }

The bosonic part of the model \rf{2}--\rf{7} is classically integrable \ci{dmv} and thus any consistent
truncation of the corresponding string equations
yields a classically integrable string model.

A 3d truncation of the metric in \rf{3}, \rf{4} is found by setting
$\z=\psi_2=0$, $\psi_1\equiv \psi$:\foot{Note that setting $\psi_1=\psi_2=0$ is not a
consistent truncation because of the $B$-field \rf{7} contribution to the $\psi_1$ equation of motion.}
\be ds^2_{A_3} = - h(\r) dt^2 + f(\r) d\r^2 + \r^2 \, d\psi^2 \ . \la{tr} \ee
The scalar curvature of this metric is
$ R= - { 2[ 3 + \k^2 - ( 3 - \k^2 ) \k^2 \r^2 + \k^4 \r^4] \ov 1- \k^2 \r^2 } $.
Similarly, from \rf{5}, \rf{6} we get ($\p_1\equiv \p$)
\be ds^2_{S_3} = \td h(r) d\vp^2 +\td f(r) dr^2 + r^2 \, d\p^2 \ . \la{tri} \ee
The $B$-field \rf{7} vanishes, i.e. we get purely metric 3d integrable models
that represent the $\k$-deformations of $AdS_3$ and $S^3$ respectively.

Reducing further by setting $\psi=0$ in \rf{tr} and $\p=0$ in \rf{tri} gives two 2d metrics
\be && ds^2_{A_2} = - { 1 +\r^2 \ov 1 - \k^2 \r^2} dt^2 + { d\r^2 \ov (1 +\r^2)( 1 - \k^2 \r^2)} \ , \la{tw} \\
&& ds^2_{S_2} = { 1- r^2 \ov 1 + \k^2 r^2} d\vp^2 + { dr^2\ov (1 - r^2)( 1 + \k^2 r^2) } \ . \la{two} \ee
These are $\k$-deformations of $AdS_2$ and $S^2$ which, like \rf{tr} and \rf{tri},
are related by an obvious analytic continuation.

Let us note that since the bosonic model based on the
deformed $S^5= SO(6)/SO(5)$ metric~\rf{5} (or deformed $AdS_5$ metric \rf{3})
can be obtained directly using the deformed coset construction of \ci{dmv1},
the same applies to their truncations to lower-dimensional $SO(n+1)/SO(n)$ cosets with $n=3,2$.
Indeed, a metric equivalent to \rf{two} was found in \ci{dmv1} as representing the deformed
$SO(3)/SO(2)$ coset, and \rf{tri} should be the metric for the similarly deformed coset $SO(4)/SO(3)$.

We shall study the deformed \ads and \adt models in detail in sections~3~and~4.
As we shall explain below, the
models based on \rf{two} and \rf{tri} are actually
not new: they are well-known deformations of the $S^2$ and $S^3$ sigma models
constructed in \ci{foz} and \ci{fat} respectively, and their classical integrability was proven earlier in \ci{luk}.

An interesting open question
is how to promote these classical integrable models to
10d conformal superstring sigma models that represent deformations of $AdS_3 \times S^3 \times T^4$
and $AdS_2 \times S^2 \times T^6$ models. This requires finding the dilaton and other fluxes that together with the
above metrics solve the 10d type II supergravity equations of motion.

\renewcommand{\theequation}{3.\arabic{equation}}
\setcounter{equation}{0}
\section{Deformed \ads model }

Let us now consider in detail the model whose bosonic part is given by the deformation
of the $AdS_3=SO(2,2)/SO(1,2)$ and $S^3=SO(4)/SO(3)$ cosets, i.e. with a metric which is the sum of \rf{tr} and \rf{tri}.

Since the two 3d metrics are related by an obvious analytic continuation
let us concentrate on the structure of the $\k$ deformation of $S^3$ in \rf{tri}, i.e.
\be ds^2_{S_3} = { dr^2\ov (1 - r^2)( 1 + \k^2 r^2) } + r^2 d\p^2 + { 1- r^2 \ov 1 + \k^2 r^2} d\vp^2 \ . \la{31} \ee
Here $\p$ and $\vp$ are two $U(1)$ isometry directions and, in addition to $U(1) \times U(1)$,
this model also has a discrete $Z_2$ symmetry (the $dr^2$ term is invariant under this change)
\be \p \leftrightarrow \vp\ , \ \ \ \ \ \ \ \ \ \ r \to \sqrt{ { 1- r^2 \ov 1 + \k^2 r^2}} \ . \la{z2}\ee
A 3d metric with exactly the same symmetries is a special case of Fateev's \ci{fat} 2-parameter renormalizable
deformation of the $SU(2)$ principal chiral model which is known to be classically integrable \ci{luk}:
as we shall explain in subsection 3.1 below, the sigma model based on \rf{31} is the same as the ``symmetric''
case of the Fateev model.

Furthermore, we shall demonstrate in Appendix A that the recently constructed 2-parameter family
of integrable Yang-Baxter deformations of the principal chiral model for group $G$ \ci{k2,k3} is equivalent to Fateev model in the
$G=SU(2)$ case and thus also contains \rf{31} as its special equal-parameter case.\foot{That the coset deformation
of \ci{dmv1} for the $SO(4)/SO(3)$ case is equivalent to the equal-parameter case of the $SU(2)$ model of \ci{k3}
was mentioned to us
by the authors of \ci{dmv}.}
Thus \rf{31}, which corresponds to the deformation of $SO(4)/SO(3)$ constructed according to \ci{dmv1},
is at the same time a special case of the Fateev model \ci{fat} and also a special case of the
bi-Yang-Baxter sigma model of \ci{k3}.

The \ads case is special compared to the \adss and \adt cosets as it also has an interpretation in terms of a
product of group spaces.
In this case the bosonic $S^3$ (and $AdS_3$) part has a {\it two}-parameter integrable deformation \ci{fat,k3},
and there is also a further deformation \ci{luk}
that includes a non-zero $B$-field coupling (i.e. a WZ term)
which we shall discuss in Appendix B.
This
gives a 3-parameter deformation of the $S^3$ model with the extra parameter being the coefficient $\qq$ of the
WZ term (with $\qq=0$ being the $S^3$ model and $\qq=1$ corresponding to the $SU(2)$ WZW model).
With the deformation parameters in the two bosonic factors identified,
there should then exist
the corresponding 2-parameter deformation of the $AdS_3 \times S^3$ supercoset model
with mixed 3-form flux discussed in \ci{kaz,htq}.
This then suggests that there should exist an extension of the $G,B$ background by the dilaton and RR fluxes
that preserves conformal invariance.
It should be noted, however, that already in the presence of the single-parameter
$\k$-deformation the dilaton becomes non-trivial
and thus S-duality transformations of the type IIB theory will change the sigma model (string-frame) metric.
Consequently, there will no longer be any symmetry between the NS-NS and R-R choices of 2-form background.\foot{Also,
S-duality need not, in general, preserve the integrability of string sigma model, cf. \ci{lm,frt}.}
For that reason it would be best to study the 2-parameter deformations of the cases $\qq=0$ and $\qq=1$ separately.
Apart from in Appendix B, here we will consider only the $\qq=0$ case, i.e. without $B$-field coupling.

Like the \adss model, the $\k$-deformed \ads model admits two special cases: $\k=\infty$ and $\k=i$.
Taking the limit $\k\to \infty$ in \rf{31} and introducing $\bar r = r^{-1}$ and $\bar \p= \k \p$ we get
$ds^2_{S_3} = \k^{-2} \big[ ( \bar r^2-1 )^{-1} d\bar r^2 + \bar r^{-2} d\bar \p^2 + { (\bar r^2 -1)} d\vp^2\big] $.
This becomes equivalent to the metric of euclidean $AdS_3$ space, i.e. the hyperboloid $H^3$ or euclidean $SL(2,R)$ group space,
after T-duality in the $\bar \p$ direction. Thus, in this sense, the $\k$-deformation relates the \ads
coset to $dS_3 \times H^3$. Interestingly, $\k=0$ and $\k=\infty$ correspond to the two (IR and UV) asymptotics
of the RG flow in the deformed $S^3$ model \ci{luk}.\foot{These are zeroes of the beta function for $\k$,
but to have a fixed point for all couplings one needs a WZ term \ci{luk}
(see also Appendix B).}

The $\k=i$ limit of the deformed \ads coset will be discussed in subsection 3.2. We shall see that the
metric takes the pp-wave form which in the light-cone gauge reduces to the sum of the complex sine-Gordon model and its analytic
continuation, which is precisely the bosonic part of the Pohlmeyer reduced theory for \ads (times $T^4$) superstring theory \ci{gt1,gt2}. Furthermore,
we shall find the dilaton and RR 3-form flux which promote, as in \ci{ mm,rt,bs}, this metric to a type IIB supergravity solution.
The fermionic part of the corresponding superstring Lagrangian is then found to match the fermionic part of the
PR theory for the \ads superstring \ci{gt2}.
Thus the $\k=i$ limit
of the deformed \ads supercoset should represent an exact embedding of the PR model of the
undeformed ($\k=0$) supercoset into string theory.

\subsection{$\k$-deformed $S^3$ as the ``symmetric'' case of the Fateev model }

Ref.~\cite{fat} proposed a two-parameter deformation the $O(4)$ sigma model that (i) preserves $U(1)\times U(1)$
symmetry and (ii) is
perturbatively renormalizable, { i.e.} that the change of the sigma model metric under the
RG flow (a shift by its Ricci tensor at one-loop order)
can be represented by a change of the two deformation parameters (and the overall
$S^3$ ``radius'' coupling constant).
The Lagrangian for the field $g\in SU(2)$ is \ci{fat}
\be
{L}_{S3} = \frac{1 }{2\big[ (1+\ll)(1+\rr) - \ll \rr M^2 \big]} \eta^{ij} \Big( \ha \Tr[\partial_i g \partial_j g^{-1}] + \ll\, L^3_i L^3_j
+ \rr\, R^3_i R^3_j \Big)
\label{Fa} \ ,
\ee
where $i,j=1,2$
and ($\s^a$ are Pauli matrices)
\be
M = \frac{1}{2} \Tr[g \sigma^3 g^{-1}\sigma^3]
~,\qquad
L^a_i = \frac{1}{2i} \Tr[\del_ig g^{-1}\sigma^a]
~,\qquad
R^a_i = \frac{1}{2i} \Tr[g^{-1} \del_i g \sigma^a]\ .
\label{defs}
\ee
In \rf{Fa} $\ll $ and $ \rr$ are two independent deformation parameters, and
we shall also use\foot{We have extracted the overall scale so that the other
the parameters used in \ci{fat} are
$u=1$, \ $a^2=(1 + \ll) ( 1 + \rr), \ b^2=\ll \rr$. }
\be \rd \equiv \ha(\ll + \rr ) \ , \ \ \ \ \ \ \ \ \rc\equiv \ha( \ll-\rr) \ . \la{39}\ee
One may
parametrize the $SU(2)$ field as
\be &&g=n_4 \id +i n_a \sigma^a\ , \ \ \ \ \ \ \ \ \qquad n_a n_a + n_4^2=1 \ , \la{3n} \\
&&n_1 + i n_2 = w\, e^{i \vvp_1} \ , \ \ \ \ \ \ \, \qquad n_3 + i n_4 = \sqrt{ 1-w^2 }\, e^{i \vvp_2} \ , \ \ \ \la{nn} \\
&& z\equiv n_1^2 + n_2 ^2 - n_3 ^2 - n_4^2\ , \ \ \ \ \ \ w^2= \ha (1 + z) \ . \la{nnn}
\ee
Then the 3d target space metric for the two-parameter model \rf{Fa}
becomes \cite{fat}
\be
\label{Fm}
ds^2_3 &=& U(z) dz^2 + D(z) d \vvp_1^2+D(-z)d \vvp_2^2 + 2 C(z) d \vvp_1d \vvp_2\ , \la{33}
\\
U(z) &=& (1-z^2)^{-1} Q(z) \ , \ \ \ \ \ \ \ \ Q\equiv {\te { 1 \ov 4}} \big[(1+\rd)^2-\rc^2 - (\rd^2-\rc^2)z^2\big]^{-1} \ ,\la{333} \\
D(z) &=& 2(1+z)\big[ 1+\rd(1+z)\big] Q(z)
\ , \qquad \ \ \ \ \
C(z) = 2\cc\ (1-z^2) Q(z)
\ . \la{35}\ \ \ \
\ee
As we shall show in Appendix A,
this model is equivalent to the $SU(2)$ case of the two-parameter integrable deformation of the principal chiral model
constructed in \cite{k2,k3}. On the one hand, this gives a simpler demonstration of
the integrability of the sigma model \rf{33} than that~in~\ci{luk}, and on the other, it
proves the renormalizability of the 2-parameter model of \cite{k2,k3} (checked in the 1-parameter, $\rr=0$, case in \ci{sq}).

There are two obvious special 1-parameter cases: left-right asymmetric ($\ll =0 $ or $\rr=0$)
and left-right symmetric ($\ll=\rr$).
For $\rr=0$ we get $\rd=\rc=\ha \ll$ and the metric \rf{33} becomes that of the squashed $S^3$
corresponding to the anisotropic $SU(2)$ chiral model \ci{che}.

In the case of left-right symmetric deformation\foot{Another special 1-parameter case is
$\ll=-\rr=\cc$, $\dd=0$ so that
$Q(z)= {1 \ov 4 } ( 1-\cc^2 + \cc^2 z^2)^{-1} $,
$ U(z)= ( 1 - z^2)^{-1} Q$,
$D(z) = 2 ( 1 + z)^{-1} Q\ ,$ $ C(z) = 2 \cc (1-z^2) Q \ . $
Performing T-dualities along $\chi_1$ and $\chi_2$
we get the metric
$ds^2 =\frac{dz^2}{4(1-z^2)(1-\cc^2+\cc^2 z^2)}
+\frac{2}{1+z}d\td \chi_1^2+\frac{2}{1-z}d\td \chi_2^2 -4\cc d\td \chi_1d\td \chi_2$
and zero $B$-field coupling.}
\be \ell= \rr =\rd\ , \ \ \ \ \ \ \rc=0 \ ,\la{syy} \ee
the metric \rf{33} simplifies to
\be \label{le}
ds^2 &=& U(z) dz^2 + D(z) d \vvp_1^2+D(-z)d \vvp_2^2 \ ,
\\
U(z) &=& \frac{1}{4(1-z^2)\big[(1+\rd)^2 - \rd^2z^2\big]} \ ,
\qquad
D(z) = \frac{1+z}{2\big[(1+\rd) - \rd z\big]}
\ .
\la{el}
\ee
This metric is manifestly invariant under the $Z_2$ symmetry: $z\rightarrow -z$ and $ \vvp_1\leftrightarrow \vvp_2$.
It is indeed equivalent to \rf{31} under the following identification of the coordinates and parameters:
\be \la{37}
r^2 = \frac{1 + z}{2\big[(1 + \rd) - \rd z\big]} \ , \ \ \ \ \ \p= \vvp_1 \ , \ \ \ \ \ \ \vp=\vvp_2 \ , \ \ \ \ \ \ \ \ \
\k^2 = 4 \rd (1 +\rd) \ . \ee
Note that $r^2= \ha ( 1 + z) + O(\k^2)$, i.e. in the undeformed case $r$ is same as $w$ in \rf{nn} (cf. \rf{nnn}).

Let us note that the metric \rf{33} becomes flat for $\rd=- \ha$, i.e. $\ll + \rr=-1$. In this case we see from \rf{37} that
$\k =i$, but the change of variables $r\to z$ degenerates, so there is no contradiction with the fact that the metric \rf{31}
remains non-trivial for $\k=i$: rescaling $z$ in \rf{33} together with taking $\rd \to -\ha$ gives the metric in \rf{31}.

As was mentioned above, there should exist an integrable 2-parameter deformation of the \ads supercoset
model whose bosonic part is given by the direct product of the 2-parameter deformed $S^3$ \rf{33} and the similarly
deformed (with the same parameters) $AdS_3$.
To be sure that this deformation will represent a conformal model one needs to find the corresponding dilaton and
type IIB fluxes promoting such a 3+3 dimensional metric to an exact type IIB supergravity solution.
Finding such a solution at the moment is an open problem even in the case of the symmetric 1-parameter deformation.

\subsection{$\k=i$ limit: equivalence to the PR model for the \ads superstring }

To shed light on the underlying type IIB background and to establish a relation to the PR model for the \ads superstring, let us now consider
the special $\k=i$ limit of the deformed \ads model \rf{tr}, \rf{tri}.
Directly setting $\k=i$ in this deformed $AdS_3 \times S^3$ metric leads to the
decoupling of the coordinates $t$ and $\varphi$. Coupling this limit with a
nontrivial rescaling of these coordinates as described in eq.~\eqref{lim}
leads to a non-trivial pp-wave metric which is a truncation of \rf{k} (we set $s=1$)
\be
&ds^2 = 4dx^+dx^- - \big[V(\alpha) + V(\beta)\big]\,dx^+{}^2
+d\a^2 + \tan^2\a\, d\psi^2 + d\b^2 + \tanh^2\b\, d\phi^2\ ,\label{PRv}
\\
& V(\a) = \sin^2 \a \ , \quad V(\b) = \sinh^2\b \ , \qquad \rho \equiv \tan\a \ , \quad r \equiv \tanh \b\ .\no
\ee
Fixing light-cone gauge,
$x^+ = \mu \tau$,
we find the following
Lagrangian
\be\la{lac}
L_{l.c.} = - (\partial_i \a \partial^i \a + \tan^2\a\, \partial_i\psi\partial^i \psi + \mu^2 \sin^2\a)
- (\partial_i \b \partial^i \b + \tanh^2\b\, \partial_i\phi\partial^i \phi +\mu^2 \sinh^2\b) ,\ \
\ee
which is precisely the bosonic part of the PR Lagrangian
for strings moving on $AdS_3 \times S^3$~\ci{gt1,gt2}.
Note that as in the \adss case, the roles of the $AdS_n$ and
$S^n$ spaces appear to be interchanged, i.e.
the $\k\to i$ limit of the deformed $AdS_3$ metric gives the PR
of the string on $R \times S^3$ (after fixing light-cone gauge), and vice versa -- the $\k \to i$
limit of the deformed $S^3$ leads to the PR of the string on $AdS_3 \times S^1$.

Let us now extend the direct product of the pp-wave space \eqref{PRv} and a torus
$T^4$ to a full solution of type IIB supergravity
by finding the corresponding dilaton and 3-form RR flux
that solve the equations (we shall assume that all other fluxes vanish)\foot{Note that since $F_3$ will be assumed to be non-zero
only in 6d subspace, its stress tensor is traceless
(or, equivalently, $F{}^{\eta\rho\sigma}F_{\eta\rho\sigma}=0$) and thus $R + 2 \nabla^2 \Phi=0$.
Then the dilaton equation may be written as $\nabla^2 e^{-2\Phi}=0$.}
\be
R +4 \nabla^\mu\nabla_\mu\Phi -4 \nabla^\mu\Phi \nabla_\mu\Phi &=& 0\la{d} \ ,
\\
\te R_{\mu\nu} + 2\nabla_\mu \nabla_\nu\Phi &=&\te \frac{1}{4}e^{2\Phi}\big(F_{\mu}{}^{\rho\sigma}F_{\nu\rho\sigma} -\frac{1}{6}g_{\mu\nu}
F{}^{\eta\rho\sigma}F_{\eta\rho\sigma} \big)\ ,
\label{Eeq}
\\
\partial_{[\mu}F_{\nu\eta\rho]}&=&0\ ,
\qquad\quad
\partial_{\mu}(\sqrt{-g}F^{\mu \nu\rho})=0\ .
\la{f}
\ee
Let us change the coordinates to complex $u,w$
so that \eqref{PRv} becomes
\be
&&ds^2 =
4dx^+dx^- - (|u|^2 + |w |^2)\,dx^+{}^2
+\frac{du\,d{\bar u}}{1-|u|^2}+\frac{dw\,d{\bar w}}{1+|w|^2}
\ ,
\label{PRme}\\
&& \qquad u = \sin \a \; e^{i\psi}\ , \ \ \ \ \ \ \ \ \ \ w = \sinh \b \; e^{i\phi} \ . \la{uw}
\ee
The Ricci scalar of this metric is non-zero because of the curved transverse space
\be
R
=\frac{4}{1 + |w|^2}-\frac{4}{1 - |u|^2}\ . \la{rer}
\ee
The solution to the dilaton equation can be written as
\be
\Phi &=& -\ln f(|u|^2)-\ln g(|w|^2) \ , \\
f(x) &=&\sqrt{1- x}\left[c_1 P_v (-1+2 x)+c_2 Q_{v}(-1+2 x)\right]\ , \ \ \ \ \ v={\ha ( c_0-1)}\ , \no
\\
g(x)&=&\sqrt{1+ x}\left[c_3 P_v (1+2 x)+ c_4 Q_v (1+2 x)\right] \ . \la{qp}
\ee
Here $c_n $ are integration constants, $c_0$ is the separation constant appearing when
splitting equation \rf{d} as
$E(u)=c_0, \ E(w)=-c_0$.
$P_v$ and $Q_v$ are the two independent solutions of the Legendre equation, where
$Q_v$ has logarithmic singularities and can thus be ignored.
$P_v$ is a polynomial if the index $v$ is an integer. The simplest choice is $v=0$ ($c_0
= 1$) when $P_0= \text{const}$ and thus
\be
\te \Phi =\Phi_0 -\frac12\ln(1- |u|^2)-\frac12\ln(1+ |w|^2) \ .\la{dii}
\ee
The transverse metric in \rf{PRme} and the dilaton \rf{dii} represent
of course the direct sum of the familiar 2d black hole
($SL(2,R)/U(1)$) solution and its analytic continuation.
It follows from \rf{dii} that $R_{\mu\nu} + 2\nabla_\mu \nabla_\nu\Phi$ has only one non-zero component
\be
R_{++} + 2 \nabla_+ \nabla_+ \Phi = 4 (1 - |u|^2 + |w|^2)\ .
\ee
It is readily seen that it can
be balanced by the solution for the RR 3-form flux $F_{\m\n\r}$ with the following real potential\footnote{We use the following conventions:
$F_3{}_{\mu\nu\rho} = (dC_2)_{\mu\nu\rho} = \partial_\mu C_2{}_{\nu\rho} + \partial_\nu C_2{}_{\rho\mu} + \partial_\rho C_2{}_{\mu\nu}\, ,$
$C_2 = \frac12 C_2{}_{\mu\nu} dx^\mu \wedge dx^\nu \, ,$ $F_3 = \frac16 F_3{}_{\mu\nu\rho}dx^\mu \wedge d x^\nu \wedge dx^\rho \, .$}
\be
C_2 = i{\te {\frac{1}{ \sqrt2}}} e^{- \Phi_0}
\big[ \cos \gamma \, (1+ |w|^2)( u d{\bar u}- \bar u du ) + \sin \gamma \, (1-
|u|^2)(w d{\bar w}- \bar w dw )\big] \wedge dx^+ \ , \la{c2}
\ee
where $\g$ is a free parameter. Motivated by matching the Pohlmeyer reduction of the $AdS_3 \times S^3$ string \ci{gt2},
which has a formal $Z_2$ symmetry interchanging the $AdS_3$ and $S^3$ parts,
a natural choice for $\g$ is $ \frac{\pi}{4}$, i.e.
\be
\cos\g = \sin \g =\te {1 \ov \sqrt 2} \ .
\ee
The above pp-wave type background $M^6 \times T^4$ thus represents the embedding of the
direct sum of the complex sine-Gordon model and its analytic continuation \rf{lac} into 10d type IIB
string theory and thus belongs to the class of models discussed in \ci{mm,rt,bs}.\foot{The embedding of this
integrable model considered in \ci{bs} used a particular $F_5$ background instead of the $F_3$ background considered here.}

It is straightforward to find the quadratic fermionic term in the GS superstring action corresponding to the above
pp-wave background
\be
&& L_{F2}= {i }
(\eta^{ij}\delta^{IJ} +\epsilon^{ij}{\s}^{IJ}_3) {\bar\theta}^I \slashed{e}{}_i \DD^{JK}_j \theta^K \ ,\ \ \ \ \ \ \slashed{e}{}_i = \Gamma_A e^A_M(X) \del_i X^M\ , \la{qq}\\
&&
\DD^{JK}_i = \del_i X^M \DD^{JK}_M \ ,\ \ \ \ \
\ \DD^{JK}_M = (\partial_M + \frac{1}{4} \omega_{M}{}^{AB} \Gamma_{AB})\delta^{JK} + \frac{1}{8} e^{\phi} \, \slashed{\rm F}_{(3)} \Gamma_M\s^{JK}_1 \ .
\ee
Here $\slashed{\rm F}_{(n)}=\frac{1}{n!}F_{A_1\dots A_n}\Gamma^{A_1\dots A_n}$
and we assume conformal gauge. The fermions $\theta^I$
($I=1,2$) are the two IIB Majorana-Weyl spinors,
and
$\s^{IJ}_1,\ \s^{IJ}_3 $ are Pauli matrices.\foot{We
use the metric $\eta_{+-}=-\tfrac12\,,\, \eta_{2 {\bar 2}}=\eta_{4 {\bar 4}}=\tfrac12\,,\, \eta_{ij}=\delta_{ij}$
and
$ e^+ = dx^+ \, , \, e^- = 4dx^- + (|u|^2+ |w|^2) dx^+ \,,$ $e^i = dz^i\,, \,
e^2 =\frac{du}{\sqrt{1-|u|^2}} \,,
\,
e^{\bar 2}=\frac{d{\bar u}}{\sqrt{1-|u|^2}}\,, \,
e^4=\frac{dw}{\sqrt{1+|w|^2}}\,,
\,
e^{\bar 4}=\frac{d{\bar w}}{\sqrt{1+|w|^2}}\,.$
The non-zero components of the spin connection are
$
\omega_{{\bar 2} 2} =-\omega_{2{\bar 2}} = \frac{1}{4}\frac{ud{\bar u} - {\bar u}du}{1-|u|^2}\,,\,
\omega_{{\bar 4} 4} =-\omega_{4{\bar 4}} = -\frac{1}{4}\frac{wd{\bar w} - {\bar w}dw}{1+|w|^2}
$.}
The explicit form of the product of the $F_3$ form corresponding to \rf{c2}
and the dilaton factor that enters the covariant derivative is
\be e^{\Phi} F_3
=
i \Big[\, \sqrt{1+ |w|^2} \, \sqrt{1-|u|^2} \big( e^2 \wedge e^{\bar 2}
+ e^4 \wedge e^{\bar 4}\big)
+ {\bar u}{ \bar w} \, e^2 \wedge e^4 - {u}{ w}\, e^{\bar 2} \wedge e^{\bar 4}\Big]\wedge e^+ \ . \la{FF}
\ee
Fixing the light-cone gauge
\be
x^+=\mu\tau \ , \qquad \Gamma^+\theta^I = 0 \ ,
\label{lc}
\ee
and rescaling the fermions by $\frac1{\sqrt \mu}$ we find ($\partial_\pm = \frac{1}{2}(\partial_0\pm \partial_1)$)
\be
L_{F2}
&=& i{\bar\theta}^1 \Gamma^- \Big[\partial_-
+ \frac{1}{8} \frac{u\partial_-{\bar u} - {\bar u}\partial_- u}{1-|u|^2} \Gamma^{2 {\bar 2}}
- \frac{1}{8} \frac{w\partial_-{\bar w} - {\bar w}\partial_- w}{1+|w|^2} \Gamma^{4 {\bar 4}}\Big] \theta^1
\cr
&& + i{\bar\theta}^2 \Gamma^- \Big[\partial_+
+ \frac{1}{8} \frac{u\partial_+{\bar u} - {\bar u}\partial_+ u}{1-|u|^2} \Gamma^{2 {\bar 2}}
- \frac{1}{8} \frac{w\partial_+{\bar w} - {\bar w}\partial_+ w}{1+|w|^2} \Gamma^{4 {\bar 4}}\Big] \theta^2
\\
&& +{\te \frac{1 }{8}}\mu \,
{\bar\theta}^I \Gamma^-
\Big[ \, \sqrt{1+ |w|^2} \, \sqrt{1-|u|^2} \big( \Gamma^{2\bar 2}+\Gamma^{4\bar4}\big)
+\ {\bar u}{ \bar w} \, \Gamma^{24} - {u}{ w}\, \Gamma^{\bar 2\bar 4} \Big]
\s_1^{IK} \theta^K\ , \no
\ee
where we have used that $\{\Gamma^+,\Gamma^-\}=-2\eta^{+-}=4$.
Returning to the original coordinates in \rf{lac} (cf. \rf{uw})
this can be rewritten as\foot{We use that
$\Gamma^{2 {\bar 2}} = \Gamma^{2+i3, 2-i3} = -2i\Gamma^{23}$,
\ $\Gamma^{4 {\bar 4}} = \Gamma^{4+i5, 4-i5} = -2i\Gamma^{45} $,
\ $\Gamma^{2 4} = \Gamma^{2+i3, 4+i5} = \Gamma^{24}-\Gamma^{35} \ +i\Gamma^{25}\ +i\Gamma^{34}$, etc.}
\be
L_{F2}&=& i{\bar\theta}^1 \Gamma^- \Big[\partial_-
- \tfrac12 \tan^2 \a\, \partial_- \psi_1 \Gamma^{23} + \tfrac 12 \tanh^2 \b\, \partial_-
\phi_1 \Gamma^{45}\Big] \theta^1
\cr
& &+ i{\bar\theta}^2 \Gamma^- \Big[\partial_+
- \tfrac12 \tan^2 \a\, \partial_+ \psi_1 \Gamma^{23}
+ \tfrac12 \tanh^2 \b\, \partial_+ \phi_1 \Gamma^{45}\Big] \theta^2
\cr
&&
-{\te \frac{i}{4}} \mu {\bar\theta}^I \Gamma^-
\Big[\cos \a\, \cosh \b \big(\Gamma^{23}+ \Gamma^{45}\big) \la{2fff}
\\ && \quad -\sin \a\, \sinh \b \big[\cos(\psi_1+\phi_1)( \Gamma^{25}+\Gamma^{34}) - \sin(\psi_1 +\phi_1)( \Gamma^{24}-\Gamma^{35})\big]\no
\Big]
\s_1^{IK} \theta^K\ .
\ee
Since the bosonic part of the light-cone gauge action
is precisely the bosonic part of the action of the Pohlmeyer reduction of the \ads superstring \cite{gt2},
it is natural to expect that the fermionic parts also match. To compare \rf{2fff}
with quadratic fermionic term in the PR action in \ci{gt2}
let us first decompose the fermions as
\be
\theta^I = \theta^I_{||} +\theta^I_\perp
~,\qquad
(1-\Gamma_{2345})\theta^I_{\perp} =0
~,\qquad
(1+\Gamma_{2345})\theta^I_{||} =0 \ .
\label{G2345projector}
\ee
$\theta^I_{||} $ and $ \theta^I_\perp$ are decoupled at the quadratic level and we find
\be
&&L_{F2}(\theta_\perp,\theta_{||}) = L_{F2}(\theta_\perp)
+ i{\bar\theta}_{||}^1 \Gamma^- \Big[\partial_-
- \tfrac12 (\tan^2 \a\, \partial_- \psi_1 - \tanh^2 \b\, \partial_- \phi_1 ) \Gamma^{23} \Big] \theta_{||}^1
\no \\
&&\qquad \qquad \qquad \qquad \qquad \! +\
i{\bar\theta}_{||}^2 \Gamma^- \Big[\partial_+
- \tfrac12 (\tan^2 \a\, \partial_+ \psi_1 - \tanh^2 \b\, \partial_+ \phi_1) \Gamma^{23}\Big] \theta_{||}^2
\la{pee}\\
&&\ \ \ \ \ \ - {\te \frac{i}{2}} \mu {\bar\theta}^I_{||} \Gamma^-
\Big[\cos \a\, \cosh \b \Gamma^{23} \la{2ff}
-\sin \a\, \sinh \b \big[\cos(\psi_1+\phi_1) \Gamma^{25} - \sin(\psi_1 +\phi_1) \Gamma^{24} \big] \Big]
\s_1^{IK} \theta^K_{||}\ .
\no
\ee
To make contact with \cite{gt2} we should choose a particular representation for the 10d Dirac matrices and also a
solution to the light-cone gauge condition
\eqref{lc} and the orthogonal decomposition \eqref{G2345projector}
(different choices may lead to different identifications of the components of $\theta^I_m$, $m= 1, ..., 32$,
and the fermions in \cite{gt2}).
In the Majorana representation of the 10d Dirac matrices and with left-handed fermions
one finds that only $\theta^I_m$ with $m={29, 30, 31, 32}$ survive the various projections and the action becomes:
\def\nquad{\!\!\!}
\be
L_{F2}=L_{F2}(\theta_\perp)
&\nquad+& \nquad 8 i\, \Big\{
\theta^1_{29}\partial_+ \theta^1_{29} +\theta^1_{30}\partial_+ \theta^1_{30}+\theta^1_{31}\partial_+ \theta^1_{31}
+ \theta^1_{32}\partial_+ \theta^1_{32} \cr & & \qquad \quad+ \,
\theta^2_{29}\partial_- \theta^2_{29} +\theta^2_{30}\partial_- \theta^2_{30}+\theta^2_{31}\partial_- \theta^2_{31}+\theta^2_{32}\partial_- \theta^2_{32} \vphantom{\Big[}
\cr
&& \quad - \tan^2 \a \,\big[
\partial_- \psi_1\,(\theta^1_{29} \theta^1_{30} + \theta^1_{31} \theta^1_{32})
+ \partial_+ \psi_1\,(\theta^2_{29} \theta^2_{30} + \theta^2_{31} \theta^2_{32}) \big] \vphantom{\Big[}
\cr
&& \quad + \tanh^2 \b\,\big[ \partial_- \phi_1\,(\theta^1_{29} \theta^1_{30} + \theta^1_{31} \theta^1_{32})
+ \partial_+ \phi_1\,(\theta^2_{29} \theta^2_{30} + \theta^2_{31} \theta^2_{32}) \big]\vphantom{\Big[}
\cr
&& \quad - \mu
\big[\cos \a\, \cosh \b \; (-\theta^2_{30}\theta^1_{29}+\theta^2_{29}\theta^1_{30}-\theta^2_{32}\theta^1_{31}+\theta^2_{31}\theta^1_{32})
\label{finalF2} \vphantom{\Big[}\\
&&\quad \qquad -\sin \a\, \sinh \b \cos(\psi_1+\phi_1)(-\theta^2_{32}\theta^1_{29}+\theta^2_{29}\theta^1_{30}-\theta^2_{31}\theta^1_{30}+\theta^2_{30}\theta^1_{31}) \vphantom{\Big[}
\cr
&&\quad \qquad - \sin \a\, \sinh \b \sin(\psi_1 +\phi_1)(+\theta^2_{31}\theta^1_{29}-\theta^2_{29}\theta^1_{31}+\theta^2_{30}\theta^1_{32}-\theta^2_{32}\theta^1_{30}) \big]\Big\}
\ . \no
\ee
Eq.~\eqref{finalF2} can then be mapped to the quadratic fermionic term in the
Lagrangian of the PR model for the \ads superstring \cite{gt2} by identifying the fields as follows\foot{\label{def_gt_partial} Here
$\a,\b,$ etc., are the fermionic fields used in \ci{gt2}. Note also that $\partial_\pm^\text{here} =\frac{1}{2}(\partial_0\pm \partial_1)=\frac{1}{2}\partial_{\pm}^{\text{there}}$.}
\be
&& \theta^1_{29}={\textstyle \frac{1}{2}}\alpha\ ,
\ \ \ \ \
\theta^1_{30}={\textstyle \frac{1}{2}}\beta\ ,
\ \ \ \ \
\theta^1_{31} ={\textstyle \frac{1}{2}} \delta\ ,
\ \ \ \ \
\theta^1_{32}={\textstyle \frac{1}{2}}\gamma \ ,
\nonumber\\[2pt]
&&
\theta^2_{29}=-{\textstyle \frac{1}{2}}\sigma\ ,
\ \ \
\theta^2_{30}={\textstyle \frac{1}{2}}\rho\ ,
\ \ \ \ \
\theta^2_{31}={\textstyle \frac{1}{2}}\lambda\ ,
\ \ \ \ \
\theta^2_{32}=-{\textstyle \frac{1}{2}}\nu\ .
\ee
The full light-cone gauge GS action for the background \rf{PRv}, \rf{dii}, \rf{c2} also contains quartic fermionic
terms (whose presence is related to the curvature of the transverse space), and there is no
doubt that they should also match the quartic fermionic terms in the PR action~\ci{gt2}.

We conclude that the GS string with the $\k=i$ metric as bosonic part, which should be the $\k=i$ limit of the deformed
\ads supercoset action (see footnote~\ref{footnoteGS}), should represent
the embedding of the ``massive'' Pohlmeyer reduced model for the undeformed ($\k=0$)
\ads supercoset into superstring theory.

Let us note that, unlike the pp-wave solutions with flat transverse space,
the type IIB solution \rf{PRv}, \rf{dii}, \rf{c2} does not have residual target-space supersymmetry
(the same is true also for the $F_5$-flux supported background
in \ci{bs}).
One may nevertheless expect the existence of hidden $(4,4)$ world-sheet supersymmetry
in the corresponding superstring theory \ci{bs}. This
parallels the discussions in \ci{gt1,gt2,goi,ht2,hom,sh} where it was argued that the PR theory for the \ads superstring
should have hidden supersymmetry.

\renewcommand{\theequation}{4.\arabic{equation}}
\setcounter{equation}{0}
\section{Deformed $AdS_2 \times S^2$ model }

Let us now consider the deformation of the \adt coset with the bosonic part given by the sigma model
corresponding to the metrics \rf{tw} and \rf{two}, i.e.
\be
&& L = -\frac{ 1+\rho^2 }{1-\k^2\rho^2} (\del_i t)^2 +\frac{1 }{(1-\k^2\rho^2)(1+\rho^2)} (\del_i \r)^2
\nonumber\\[1pt]
&& \qquad +
\frac{1-r^2}{1+\k^2 r ^2} (\del_i \varphi)^2 +\frac{1}{(1+\k^2 r^2)(1- r^2)} (\del_i r)^2 \ . \la{41}
\ee
The 1-parameter deformed $S^2$ model appearing here was first considered in \ci{foz} (with its integrability shown in \ci{luk})
and then rederived as a special case of the 1-parameter coset deformation construction in \ci{dmv1}.

In the special limit $\k\to \infty$ the Lagrangian \rf{41} directly reduces (up to overall $\k^{-2} $ factor)
to that of the 2d de Sitter plus 2d hyperboloid ($dS_2 \times H^2$)
model (without an additional T-duality required in \adss case in section 2.1).
In the $\k=i$ limit \rf{41} the target space metric becomes flat;
combining this limit with an additional rescaling of coordinates as
in \rf{lim} then leads to a 4d pp-wave model which is a truncation of \rf{k} or \rf{PRv}.

Below we shall first show how \rf{41} emerges as the bosonic part of the deformed \adt supercoset action constructed
using the same method as in \ci{dmv} and then discuss the special case of $\k=i$.

\def\diag{\operatorname{diag}}
\def\dpm{\partial_\pm}
\def\dpl{\partial_+}
\def\dm{\partial_-}
\subsection{Deformed supercoset Lagrangian }

The superstring theory in $AdS_2 \times S^2 \times T^6$ is closely related \ci{sor}
to the GS model based on the supercoset~\ci{beb}
\be
{PSU(1,1|2)\ov SO(1,1)\times U(1)}
\label{supercoset}
\ee
which belongs to the class of supercosets in which the denominator is the fixed point of a $Z_4$
automorphism of the numerator.
It moreover turns out that the construction of the fermionic part of the action
is quite sensitive to the $Z_4$ action.
A possible choice for its generator that proves to be particularly useful, given in Appendix C of \cite{gt1}, is\footnote{$\omega$
can be written in the form $\omega(M) = -K^{-1}M^{st}K$, where $K = \diag(\sigma_3,\sigma_3)$ and $st$ denotes the supertranspose:
$\begin{pmatrix}
M^{(a)} & M^{(f_1)} \cr
M^{(f_2)} & M^{(s)}
\end{pmatrix}^{st} =
\begin{pmatrix}
M^{(a)}{}^t & - M^{(f_2)}{}^t \cr
M^{(f_1)}{}^t & M^{(s)}{}^t
\end{pmatrix}$. Furthermore, this implies that $\omega$ satisfies $\omega(MN) = -\omega(N)\omega(M)$.}
\be
\omega(M) =
\begin{pmatrix}
-\sigma_3 M^{(a)}{}^t\sigma_3 & \sigma_3 M^{(f_2)}{}^t\sigma_3 \cr
-\sigma_3 M^{(f_1)}{}^t\sigma_3 & -\sigma_3 M^{(s)}{}^t\sigma_3
\end{pmatrix} \ , \qquad
M =
\begin{pmatrix}
M^{(a)} & M^{(f_1)} \cr
M^{(f_2)} & M^{(s)}
\end{pmatrix} \ .
\label{bad_Z4}
\ee
This $\omega$ identifies the gauge group generators as $\diag(\sigma_1,0)$ and $\diag(0,i\sigma_1)$.

The GS Lagrangian for this supercoset may be written as\footnote{We use the normalization in which the (super)trace of squares of the bosonic
Cartan generators equals 2.
}
\be
&& {L}_0 = \pi^{ij} \, \STr[J_i \, d_0\, J_j]\ ,
\qquad
\qquad \pi^{ij} \equiv \sqrt {-g} g^{ij} - \epsilon^{ij} \ , \la{de4} \\
&& \ \ \ \ \ \
J_i =g^{-1}\partial_i g\ ,
\qquad
d_0 \equiv P_1+2P_2 -P_3 \ ,
\ee
where $P_k$ are projectors onto subspaces with eigenvalue $i^k$ under the action of the $Z_4$ automorphism.

The one-parameter $\eta$-deformation of this supercoset Lagrangian constructed according to \ci{dmv} is\foot{Recall that
in terms of $\k$ in \rf{1} we have $\eta = \k^{-1} [\sqrt{ 1 + \k^{2}} - 1 ]$.}${}^{,}$\footnote{
The action \rf{2}
corresponding to the Lagrangian in \rf{de}
is normalized as in \cite{abf}.}
\be && \label{de}
{ L} = c_\eta\, \pi^{ij}\, \STr[J_i \, d_\eta \circ \frac{1}{1-\eta R_g\circ d_\eta}\, J_j] \ , \\
&& \qquad
d_\eta \equiv P_1+ 2 c_\eta^{-1} 
P_2 -P_3\ ,
~~~~\qquad
c_\eta \equiv 1-\eta^2 \ .
\label{def}
\ee
The operator $R_g$ acts on the superalgebra as
\be
R_g(M) = g^{-1}R(g M g^{-1})g \ ,
\label{Rgaction}
\ee
where the operator $R$ multiplies the generators
corresponding to the positive roots by $-i$, those corresponding to the negative roots by $+i$
and annihilates the Cartan generators.
It is possible to choose the positive roots to be generators whose nonzero entries are above the
diagonal, which corresponds to considering the distinguished Dynkin diagram for $PSU(1,1|2)$.\foot{The
relation between the deformations corresponding to different choices of Dynkin diagram is unclear.
}

Independently of the choice of $Z_4$ automorphism, a systematic approach to expanding the Lagrangian in terms
of coordinate fields is to represent the action of the operators $d$ and $R_g$ in the adjoint representation. We
also introduce two auxiliary matrices ${\tilde A}$ and ${\hat A}$:
\be
d_\eta (T^a) = {\rm d}_\eta{}^a{}_{b} T^b\ ,
\qquad
R(T^a) = R^a{}_b T^b\ ,
\qquad
gMg^{-1} = M_a {\tilde A}^a{}_b T^b
\qquad
g^{-1}Mg = M_a {\hat A}^a{}_b T^b
\ ,
\ee
where $M$ is a generic element of the algebra of $PSU(1,1|2)$.
Then
\be\la{dee}
{ L} = c_\eta\, \pi^{ij}\, J_i{}_a\, J_{jd} \Omega^d{}_u{\rm d}_\eta{}^u{}_v {\rm g}^{av}
\qquad\qquad
\Omega^{-1} \equiv \id - \eta\, {\rm d}_\eta {\tilde A} R {\hat A} \ . 
\ee
Using that ${\rm g}^{ae}{\rm d}_\eta {}^c{}_e={\rm g}^{ce}{\rm d}_\eta {}^a{}_e$ the Lagrangian
can be written in terms of $ \Omega{}^b{}_c {\rm g}^{ce} \equiv \Omega^{be}$,
where $\Omega^{be}$ may be interpreted as a
deformation of the group-invariant metric ${\rm g}^{be}=\STr[T^bT^e]$, to which it reduces in the limit $\eta\rightarrow 0$.

We parametrize the coset elements as \cite{abf}
\be
&& g = g_B g_F \ , \qquad
\label{Bcosetrep}
g_B =
\begin{pmatrix}
g_A & 0_{2\times 2} \cr
0_{2\times 2} & g_S
\end{pmatrix} \ ,
\qquad 
g_F =\exp
\begin{pmatrix}
0_{2\times 2} & 
f_1 \cr
f_2 & 0_{2\times 2}
\end{pmatrix}\ ,
\\ &&
g_A = e^{{1 \ov 2} it\sigma_3}
\begin{pmatrix}
\rho_+ & i \rho_- \cr
-i\rho_- & \rho_+
\end{pmatrix}\ ,
\qquad\ \ \ \
\rho_\pm = \frac{1}{\sqrt{2}}\sqrt{\sqrt{\rho^2+1}\pm 1}\ ,
\\
&&g_S = e^{{1\ov 2 } i\varphi \sigma_3}
\begin{pmatrix}
r_+ & r_- \cr
- r_- & r_+
\end{pmatrix}\ ,
\qquad \qquad
r_\pm = \frac{1}{\sqrt{2}}\sqrt{1\pm \sqrt{1-r^2}}\ ,
\ee
with the fermions $f_1$ and $f_2$ related by a reality condition.
The bosonic part of the Lagrangian~\rf{dee}, found by setting
the fermions to zero, then coincides
with \rf{41}, which is a truncation of the deformed \adss model.

Using the parametrization of $g_F$ given in eq.~\eqref{fparam} of Appendix C we can explicitly construct
the part of the deformed action quadratic in fermions. The resulting expression is rather
lengthly, hence we will not present it here. However, it is useful to perform 
a simple test of one-loop UV finiteness of the deformed sigma model \rf{de4} by 
expanding to quadratic order in fields around the BMN type 
geodesic $t=\varphi = \tau$ of the deformed metric in \rf{41}.
Doing so we find the following quadratic bosonic Lagrangian in 
conformal gauge\foot{Note we have rescaled the fields to
put the Lagrangian in canonical form. Recall that here we use $\dpm = \tfrac12(\partial_0 \pm \partial_1)$.}
\begin{equation}
L = -4\dpl \tilde t \dm \tilde t + 4 \dpl \tilde \varphi \dm \tilde \varphi
 + 4 \dpl \rho \dm \rho - \frac{(1+\eta^2)^2}{(1-\eta^2)^2} \rho^2
 + 4 \dpl r \dm r - \frac{(1+\eta^2)^2}{(1-\eta^2)^2} r^2\ ,
\end{equation}
while the Lagrangian quadratic in fermions (again with a rescaling of the fields) is
\begin{equation}
L = - q_1 \dpl q_1- s_1 \dm s_1 - \frac{1+\eta^2}{1-\eta^2} q_1 s_1 - q_2 \dpl q_2 - s_2 \dm s_2 - \frac{1+\eta^2}{1-\eta^2} q_2 s_2\ .
\end{equation}
Thus, apart from the unphysical (longitudinal) 
fluctuations $\tilde t$ and $\tilde \varphi$, we get 
$2$ bosonic and $2$ fermionic excitations with the same mass $\tfrac{1+\eta^2}{1-\eta^2}=\sqrt{1 + \k^2}$.
As a result, the one-loop partition function is finite. 

\subsection{Special case of $\k=i$ \label{ads2 varkappa eq i}}

Taking the limit $\k \to i$ as in \rf{lim}, \rf{k} we get the following 4d pp-wave metric (cf. \rf{PRv})\foot{Note
that taking the limit $\k=i$ directly in \rf{41}, i.e. without the rescaling of the coordinates as in \rf{lim},
gives a flat space 4d model. In general,
the scalar
curvatures of the 2d metrics \rf{tw} and \rf{two} in \rf{41} are
$ R_A = -2 (1+ \k^2) { 1 + \k^2 \r^2 \ov 1 - \k^2 \r^2 } $ and $
R_S = 2 (1+ \k^2) { 1- \k^2 r^2 \ov 1 + \k^2 r^2 }$, i.e.
they vanish for $\k=i$. While the curvature is invariant under coordinate transformation in \rf{lim},
this vanishing is still in agreement with the fact that the resulting 4d pp-wave metric has flat transverse part.
}
\be && ds^2 = 4dx^+ dx^- - V(\a,\b) dx^+{}^2 + d\a^2 + d\b^2 \ , \ \ \ \ \ \ \ \ \ \ \\
&& V= \sin^2 \a + \sin^2 \b = | \zeta (v)|^2 \ , \ \ \ \ \ \ \zeta(v) = \sin v \ , \ \ \ \ \ \ \ v= \a + i \b \ .
\ee
As in \ci{mm,rt,bs} this pp-wave metric can be promoted to a string solution with constant dilaton
by adding a 4d vector field background (which may be viewed as an effective reduction of an RR field strength
in the 10d space $M^4 \times T^6$)
\be && A= [\zeta(v) + \bar \zeta(\bar v) ] dx^+ \ , \ \ \ \ \
F= dA = \big[\zeta'(v) dv + \bar \zeta'(\bar v) d\bar v \big] dx^+ \ . \la{faf}
\ee
This $F$ solves Maxwell's equations and we also have
$R_{++} = \ha \del_r \del_r V = F_{+ v} F_{+ \bar v}$. This background
preserves 4d space-time supersymmetry.

The resulting light-cone gauge string action has a bosonic part which is the same as the bosonic part of the
PR action for the \adt superstring model \ci{gt1,gt2}. Furthermore,
as in the \adss and \ads cases the roles of the $AdS_n$ and $S^n$ spaces
are interchanged as we interpolate from $\k=0$ to $\k=i$.

The full PR action is the same as that of the (2,2) world-sheet supersymmetric
sine-Gordon model \ci{gt1}.
An equivalent action should be found from the GS action
in the RR background \rf{faf} in the light-cone gauge (the term quartic in fermions vanishes in the light-cone gauge).
The same fermionic terms come out of the light-cone gauge fixed supercoset Lagrangian \rf{dee}
computed in the light-cone kappa-symmetry gauge. We include some details of the derivation in
Appendix~\ref{appC} where we also discuss the naive $\varkappa=i$ limit which leads to a flat space 
theory.~\foot{It appears that depending on the choice of Dynkin diagram, taking the
limit $\k\rightarrow i$ in \rf{dee} may require a rescaling of the fermionic variables.}

We conclude that
the GS string theory with bosonic part given by the deformed \adt at $\k=i$ is equivalent to
the deformed \adt supercoset model with $\k=i$ (see footnote~\ref{footnoteGS}) and represent
an effective embedding of the massive integrable Pohlmeyer reduced model for the \adt superstring
into string theory.

\section{Concluding remarks}

In this paper we explored some limits and low-dimensional analogs of the
deformed \adss supercoset integrable model constructed in \ci{dmv} with no explicit supersymmetry
but with classical quantum group symmetry.

A remarkable feature of this model is the relation \ci{abf} of the corresponding light-cone gauge S-matrix
to the real $q$ deformed S-matrix \ci{bk,hhmphase}. 
The latter also interpolates \ci{ht2,hmh,hhmphase} (for $q$ being a root of unity)
between the non-relativistic \adss ``magnon'' S-matrix and the massive relativistic S-matrix
of the Pohlmeyer reduced model for the \adss superstring.

We have studied the deformations of the low-dimensional \ads and \adt models and in these cases
made the relation to the Pohlmeyer reduced theory explicit at the Lagrangian level.
This was demonstrated by showing that in the $\k=i$ limit the deformed model reduces to a certain
pp-wave model that, in light-cone gauge, becomes equivalent to the generalized sine-Gordon model
representing the PR theory for the undeformed supercoset.
The details of a similar Lagrangian
relation in the \adss case and the issue of (non-)unitarity in the $\k=i$ limit
remain to be clarified.

We have also pointed out the possible existence of multiparameter deformations of the \ads supercoset,
clarifying the relation between the deformed $S^3$ bi-Yang-Baxter model of \ci{k2,k3} and the Fateev model \ci{fat}
which itself is a special case of the integrable 3d model found in \ci{luk}.

Among many other open questions, it would be important to understand the meaning of the deformations
suggested in \ci{k1,k2,dmv1,dmv} at a path integral level. That may help confirm that the deformed \adss
model of \ci{dmv} preserves the conformal invariance (as suggested by its classical kappa symmetry)
and thus that the corresponding target space background solves the type IIB string Weyl invariance conditions.
In fact, we have already provided several strong tests of the UV finiteness of the deformed supercoset 
model: (i) at $\k=\infty$ it is related to the finite $dS_5 \times H^5$ model; (ii) at $\k=i$ it is related to a
finite pp-wave model representing the superstring embedding of the PR model; (iii) 
in the deformed \adt case in section 4.1 we explicitly checked the one-loop UV finiteness by expanding 
near a BMN-type geodesic. It would 
nevertheless be useful to confirm the one-loop 
finiteness of the model \ci{dmv} for generic $\k$ and generic world-sheet background. 

Assuming the deformed \adss model is one-loop finite, there is still a question about higher loop orders, 
i.e. the inverse string tension $\a' \sim T_0^{-1}=g^{-1}$ corrections. While hidden higher symmetries of the model of \ci{dmv} 
may guarantee that its structure is preserved by divergent (local) loop corrections, 
to maintain it as a solution of the type IIB superstring Weyl invariance conditions one may need 
to deform the parameter $\k$ order by order in $1/g$ (starting with 4-loop $\a'^3 \sim g^{-3}$ order). 
If this happens, then the two parameters that enter the exact quantum light-cone S-matrix 
may be non-trivial functions of $\k$ and $g$ appearing in the classical string action \rf{2}.
In this case the semiclassical expression for $q$ \ci{abf} in \rf{1} may require a modification. 

While this paper was in preparation
we received \ci{holm} which discusses a similar construction to~\ci{dmv} except
with the generalized sine-Gordon model as its starting point and then interpolating to
the Hamiltonian for the light-cone gauge superstring.
Clarifying the relation between \ci{holm} and \ci{dmv} may help to understand this interpolation and the related issue of unitarity.

\section*{Acknowledgments}
We thank R. Borsato, T. Hollowood, O. Lunin, M. Magro, J. L. Miramontes, S. van Tongeren and B. Vicedo for
useful discussions.
We also thank G. Arutyunov and K. Zarembo for comments on original version of this paper. 
BH is supported by the Emmy Noether Programme ``Gauge fields from Strings" funded by the German Research Foundation (DFG).
The work of RR is supported by the US DoE under contract DE-SC0008745.
The work of AAT is supported by the ERC Advanced grant No.290456
and also by the STFC grant ST/J000353/1.

\appendix
\section{Equivalence of 2-parameter $SU(2)$ Yang-Baxter sigma model \\ to Fateev model }
\def\theequation{A.\arabic{equation}}
\setcounter{equation}{0}

It was shown in \cite{k2,k3} that the Lagrangian for $g\in G$
\be
L_K = (\eta^{ij}+\epsilon^{ij})\Tr\big[J_i\frac{1}{1-\alpha R -\beta R_g} J_j\big] \ , \qquad \qquad J_i = g^{-1}\del_ig = J_{ia} T^a\ ,
\label{Kl}
\ee
defines an integrable two-parameter ($\a,\b)$ deformation of the principal chiral model for group $G$.\foot{Let us note that this
integrable deformation
is different from the one based on a gauged WZW type construction in \ci{tse,sfet} , which is
related  \ci{sfet} to non-abelian T-duality.
Also, the parameters $\a,\b$ here should not be confused with coordinates used in the main text (cf. \rf{PRv}).}
Here the
operator $R$ acts on the generators $T^a$ as follows:
it
multiplies the generators corresponding to positive roots by $i$, the generators corresponding to negative roots by $-i$
and annihilates the Cartan subalgebra. The operator $R_g$ acts on the algebra of $G$ similarly to eq.~\eqref{Rgaction},
\be
R_g( T^a) = g^{-1}R(g T^a g^{-1})g \ , \ \ \ \ \ \ \ \ \ R(T^a) \equiv {\cal R}^a{}_b T^b \ ,
\ee
where ${\cal R}$ is the part of $R$ in eq.~\eqref{Rgaction} which acts on the generators of one $SU(2)$ factor and $\Tr[T^aT^b]=2\delta^{ab}$.
Then
\be
&& L_K= (\eta^{ij}+\epsilon^{ij})\, \Omega^{ab}(g)\, J_{ia} J_{jb} \ , \la{aa1} \\
&&\Omega^{-1} =\id -\alpha {\cal R} -\beta A(g){\cal R}A^{-1}(g) \ , \ \ \ \ \ \ \ \
gT^a g^{-1} \equiv A^a{}_b(g) T^b \ . \la{b1b}
\ee
The deformation \eqref{Kl} may thus be interpreted as
picking up a particular
nonstandard ($G$ non-invariant and in general non-symmetric) group space ``metric'' $\Omega$.

For $G=SU(2)$ generated by the Pauli matrices one finds that ${\cal R}^a{}_b$ is given by
\begin{equation}
{\cal R} = \begin{pmatrix}
0 & 1 & 0 \\
-1 & 0 & 0 \\
0 & 0 & 0
\end{pmatrix}\ .
\end{equation}
Choosing the following parametrization of the group element:
\be
g=g_3(\phi_1+\phi_2)\, g_1(r) \,g_3(\phi_1-\phi_2) \ ,
\quad
g_3(\phi) = \exp(\frac{i}{2}\phi \sigma_3)\ ,
\quad
g_1 (r) = r \id +i \sqrt{1-r^2}\, \sigma_1 \ , \la{ggg}
\ee
and using the explicit expressions for the matrices
$A(g)$ and ${\cal R}$, one finds that the symmetric and antisymmetric parts of $\Omega$ in \rf{b1b} are
\be
&&\frac{1}{2}(\Omega +\Omega^T)= \id - H^{-1} \Big[\alpha^2 {\cal R}^2 +\beta^2 A{\cal R}^2A^{-1}-\alpha\beta({\cal R}A{\cal R}A^{-1}+A{\cal R}A^{-1}{\cal R})\Big]
\ , \\
&&\frac{1}{2}(\Omega -\Omega^T)= H^{-1} \big(\alpha {\cal R} +\beta A{\cal R}^2A^{-1}\big)\ , \ \ \ \ \
H\equiv 1 + (\alpha - \beta)^2 +4 \alpha\beta r^2 \ .
\ee
The antisymmetric part of $\Omega$,
representing a WZ-type term in \rf{aa1}, contributes just a total derivative and thus may be ignored.
The model is therefore defined by
the symmetric part of $\Omega$ corresponding to the following target space metric
\be ds^2 &=&{1 \ov 1+(\alpha+\beta)^2 r^2+(\alpha-\beta)^2 (1-r^2)} \Big[ \frac{dr^2}{1-r^2}
+ r^2\left[1+(\alpha+\beta)^2 r^2\right] d\phi_1^2 \cr
&+& \left(1-r^2\right) \left[1+(\alpha-\beta)^2\left(1-r^2\right)\right] d\phi_2^2
+ 2 (\alpha^2-\beta^2) r^2 \left(1-r^2\right) d\phi_1 d\phi_2 \Big]
\ .
\label{Kme}
\ee
Using the notation as in \rf{3n}--\rf{nnn} (with $w\to r, \ \chi_1 \to \p_1, \chi_2 \to -\phi_2$)
the resulting Lagrangian may be written also in a form similar to \eqref{Fa} ($R_i^a \equiv J^a_i$)
\be
L_K = \frac{1}{2(1+ \alpha^2 + \beta^2 + 2 \alpha \beta M)}\eta^{ij} \Big[\ha \Tr(\partial_i g \partial_j g^{-1})+ (\alpha L^3_i +\beta R^3_i)(\alpha L^3_j +\beta R^3_j) \Big] \ .
\label{KF}
\ee
Here $M$ and $L^3_i,R^3_i$
are defined as in \rf{defs} but now in terms of $g$ given in eq.~\rf{ggg}.

Despite the apparent dissimilarity between the Lagrangians in \eqref{KF} and
\eqref{Fa} there exists a reparametrization that relates them, i.e.
a coordinate transformation that maps the
metric \eqref{Kme} into the one in eq.~\eqref{33}. One is to identify the parameters and the radial coordinates as follows:
\be
\alpha &=& \sqrt{(\rd-\cc) (\dd+\cc +1) }=\sqrt{ \rr (\ll +1) } \ , \quad
\qquad \b = \sqrt{(\dd+\cc) (\dd-\cc +1) }=\sqrt{ \ll (\rr +1) }
\ ,\ \ \no \\
r^2 &=& \frac{1}{2}+\frac{ (1 + 2 \dd) z - \sqrt{\big[(1 + \dd)^2 - \cc^2\big] (\dd^2 - \cc^2)}\, (1-z^2) }{2 \big[(1 + \dd)^2 - \cc^2\big]
- 2(\dd^2 - \cc^2) z^2} \ .
\ee
In the special case of $\b=0$ the matrix $\Omega$ in \rf{b1b}
becomes constant and the model reduces to the squashed 3-sphere one
\be
L_K (\b=0) = \frac{1}{1+\alpha^2}\eta^{ij} \Big[J^1_iJ^1_j + J^2_iJ^2_j + (1 + \a^2) J^3_i J^3_j\Big] \ .
\ee
In the equal-parameter case,
\be
\alpha=\beta\equiv \ha \k \ ,
\ee
the metric \eqref{Kme} becomes diagonal and is readily seen to be equivalent to the metric in \eqref{31}.
The coordinate transformation \eqref{37} then maps it to the symmetric case of Fateev model~\eqref{le}.

\def \bac {{\bar c}}
\def \bah {{\bar h}}
\def \qq {{\rm q}}
\def \ka {\kappa}
\section{4-parameter integrable 3d model with a WZ term }
\def\theequation{B.\arabic{equation}}
\setcounter{equation}{0}

Given that the 2-parameter Fateev model appears as a deformation of the $SO(4)/SO(3)$ coset
there should exist a similar 2-parameter deformation of the \ads supercoset with
bosonic part, consisting of a sum of the deformed $AdS_3$ and $S^3$ spaces, being supported by some combination of RR fluxes (and dilaton).
At the same time, there is also another deformation of the 3-sphere or $SU(2)$ principal chiral model
(and thus also of the \ads supercoset \ci{kaz,htq})
corresponding to adding a WZ term with an arbitrary coefficient $\qq$ (with $\qq=1$ as the WZW model case).
One should then expect to find an integrable 3-parameter deformation of $S^3$ (or $AdS_3$)
and thus of the \ads supercoset.

Indeed, a 4-parameter integrable deformation generalizing Fateev's 2-parameter deformed $S^3$ model
to the presence of a $B$-field coupling was constructed by Lukyanov \ci{luk}.
Below we shall review the sigma model of \ci{luk} and
suggest
that one of the two additional parameters is related to the
WZ deformation parameter $\qq$, while the other should have a ``trivial'' origin as a
T-duality (TsT or $O(2,2)$ duality) transformation parameter on the two isometric directions of the model.

The action of a 3d sigma model with two translational isometries along $(\chi_1, \chi_2)$ may be written as (cf. \rf{33})
\be
&&L= T \Big[ U(z) \del_+ z \del_- z + D(z) \del_+ \chi_1 \del_- \chi_1 + \hat D (z) \del_+ \chi_2 \del_- \chi_2 \no \\
&& \qquad\qquad +
(C+ B)(z) \del_+ \chi_1 \del_- \chi_2 + (C - B)(z) \del_+ \chi_2 \del_- \chi_1 \Big] \ , \la{b1}
\ee
where $C$ is an off-diagonal 3d metric component and $B$ is the coefficient in the 2-form $B_2= B(z) d\chi_1 \wedge d\chi_2$.
The functions in \rf{b1} have the following explicit form \ci{luk}\foot{We write the background in terms of the coordinates
$(\chi_1, \chi_2)$ related to $(u,w)$ in \ci{luk} by $ \chi_1 = \ha R^{-1} ( v - w)$, $\chi_2 = \ha ( v + w)$,
$R^2 = {(c-1)(\bac +1) \ov (c+1)(\bac -1) }$. We have absorbed an overall constant in $T$,
i.e. effectively
setting $g^2$ of \ci{luk} to 4.}
\be
&&\hspace{-15truemm}
U= { m^2 \ov 4(1-z^2) ( 1 - \ka^2 z^2) } \ , \no \\
&& \hspace{-15truemm}
D= R^2 (1+z) \big[2 + \ka ( p^2 + p^{-2}) - \ka (2 \ka + p^2 + p^{-2}) z \big]\, Q(z) \ , \ \ \ \ \no \\
\no
&&\hspace{-15truemm}
\hat D = (1- z) \big[2 + \ka ( p^2 + p^{-2}) + \ka (2 \ka + p^2 + p^{-2}) z \big]\, Q(z) \ ,\no \\
&& \hspace{-15truemm}
C= \ka (p^2 - p^{-2} ) R (1- z^2) \, Q(z) \ , \ \ \ \ \ Q(z) \equiv
{ ( c +1 ) (\bac -1) \ov 4 (1-\ka^2) ( c + z) (\bac -z) }\ , \no \\
&& \hspace{-15truemm}
B= - { m \ov c + \bar c} ( R + 1 ) \big[ \ h ( c^2 -1) (\bac -z) - \bah ( \bac^2 -1) (c + z) \big] \, Q(z)\ ,\la{b2} \ \ \ \ \
\\
&& \hspace{-2truecm} c^2\equiv { 1 + h^2 \ov \ka^2 + h^2} \ , \ \ \ \bac^2\equiv { 1 + \bah^2 \ov \ka^2 + \bah^2} \ , \ \ \
m^2 \equiv (\ka + p^2)(\ka + p^{-2} ) \ ,
\ \ \ R^2 \equiv {(c-1)(\bac +1) \ov (c+1)(\bac -1)} \ . \la{bb2}
\ee
The 4 independent parameters used in \ci{luk} are $\ka, p, h, \bah$, where $\ka\in [0,1]$ should not be confused with $\k$ in \rf{1}, \rf{4}.
In the special case of
\be h=\bah=0 \ , \ \ \ \ \ \ \ \ c=\bac = \ka^{-1} \ , \ \ \ \ \ \ R=1 \ ,\ \ \ \ \ \ \ Q(z) = {1 \ov 4(1 - \ka^2 z^2) } \ , \la{b4} \ee
the $B$-field vanishes and this model reduces \ci{luk} to the Fateev model \rf{33}
with the following identification of parameters
\be && \rd =\ha (\ll + \rr) = - \ha \ka m^{-2} ( 2 \ka + p^2 + p^{-2}) \ , \ \ \ \
\rc =\ha (\ll - \rr) = \ha \ka m^{-2} ( p^2 - p^{-2}) \ , \ \la{b5}\\
&& a^2 = (1+ \ll) (1+ \rr)= m^{-2} \ , \ \ \ \ \ \ b^2 = \ll \rr = \ka^2 m^{-2} \ , \ \ \ \ \\
&&\ll= - { \ka \ov \ka + p^2} \ , \ \ \ \rr= - { \ka \ov \ka + p^{-2}} \ , \ \ \ \ka^2= { \ll \rr \ov (\ll+1) (\rr+1) } \ , \ \ \
m^2 = { 1 \ov (\ll+1) (\rr+1) } \ , \la{kaa}
\ee
with $\hat D(z) = D(-z) $.
In the 1-parameter deformation case $\ll=\rr$ corresponding to $p=1$
(see \rf{syy}, \rf{37}) we have $\k^2= 4 \ll ( \ll+1)$ while $\ka^2 = { \ll^2 \ov (\ll+1)^2} $, i.e.
\be
\ll =\rr= { \ka \ov 1-\ka } \ , \ \ \ \ \ \ \ \ \ \ \ \ \ \ \ \ \k = { 2 \sqrt{ \ka} \ov 1-\ka } \ . \la{kaa1}
\ee
Thus $\ka= \eta^2$ where $\eta$ is the deformation parameter in \rf{1}.

A special case with a non-zero WZ term is found for $\ka=0$:
\be
&&\ka=0 \ , \ \ \ \ m=1 \ , \ \ \ \ h=\bah \ , \ \ \ \ c= \bac= (1 + h^{-2})^{1/2} \ , \ \ \ \
R=1 \ ,\la{b6} \\
&&U= {1 \ov 4(1-z^2)} \ , \ \ \ \
Q= { 1 \ov 4[1 + h^2 (1- z^2)]} \ , \no
\\ && D(z)=\hat D(-z) = 2 (1+z) Q(z)\ , \ \ \ \ C=0 \ , \ \ \ \
B= { 2 \ov \sqrt{ 1 + h^2} } \, z\, Q(z) \ .
\ee
This background represents a familiar marginal deformation (with parameter $h$) of the $SU(2)$ WZW model:
\be && ds^2 = d \theta^2 + { 1 \ov 1 + h^2 \sin^2 2 \theta } \big( \cos^2 \theta \, d\chi_1^2
+ \sin^2 \theta \, d\chi_2^2\big) \ , \no\\
&&
B_2 = { 1 \ov 2 \sqrt{ 1 + h^2 } } \
{ \cos 2 \theta\ov 1 + h^2 \sin^2 2 \theta } d \chi_1 \wedge d \chi_2 \ . \la{b12}
\ee
It can be constructed by starting with the gauged WZW model for $SU(2) \times U(1) /U(1)$
or by applying an $O(2,2)$ T-duality transformation
to the $SU(2)$ WZW model (see, e.g., \ci{hoh,has}).

It is possible to make the $\ka\to 0$ limit more non-trivial by setting \ci{luk}
\be
&&\ka \to 0\ , \ \ \ \ \ p^2 ={ \ka \ov m^2-1}
\to 0
\ , \ \ \ \ \ \ \ \ \ \ \ \ \ h=\bah = {\ka \, \qq \ov \sqrt{1-\qq^2}}
\to 0 \ , \la{b8}
\\
&&
c=\bac \to
\ka^{-1} \sqrt{1 -\qq^2} \to \infty \ , \ \ \ \ R\to 1 \ , \ \ \ \ \ \ m, \qq = {\rm fixed} \ ,
\\
&& U=\fo {m^2 \ov1-z^2} \ , \ \ \ \
Q= \fo \ , \ \ \ D(z)= \hat D(-z) = \fo ( 1 + z) [2 + (m^2 -1) ( 1-z) ] \ , \ \ \ \ \\ &&
C= - \fo (m^2 -1) \, ( 1 - z^2) \ , \ \ \ \qquad
B= \ha m \qq\, z \ ,
\ee
where
$m$ and $\qq$ are the remaining fixed parameters related to the squashing of $S^3$ and the WZ term coefficient respectively.
The resulting squashed $S^3$ metric and $B$-field are
\be
&& ds^2 = m^2 d \theta^2 + \cos^2 \theta \big[1 + (m^2 -1) \sin^2 \theta \big] d\chi_1^2
+ \sin^2 \theta \big[1 + (m^2 -1) \cos^2 \theta \big] d\chi_2^2\no\\
&&\qquad \qquad\qquad -\ (m^2 -1) \sin^2 \theta \, \cos^2 \theta\, d \chi_1 d \chi_2 \ , \ \ \ \ \ \ \ \ \ \ z=\cos 2 \theta \ , \\
&& \ \ \
\ \ \ \ B_2 = \ha m \, \qq \, \cos 2 \theta\, d \chi_1 \wedge d \chi_2 \ . \la{b10}
\ee
For $m=1$ the corresponding Lagrangian becomes that of the
$SU(2)$ principal chiral model with a WZ term with coefficient $\qq $ ($\qq=1$ is the case of the WZW model).

Another special case is $\ka=1$ (or $\k=\infty$, cf. \rf{kaa},\rf{kaa1}) when after some
parameter and coordinate redefinitions \ci{luk}
the background becomes equivalent to that of the marginal deformation of the euclidean $SL(2,R)$ WZW model \ci{hoh,hort,for}.

Like the Fateev model \ci{fat}, the above 4-parameter model is renormalizable \ci{luk},\foot{This was checked \ci{luk}
only in one-loop approximation.
However, since the model has 3d target space, the corresponding curvature tensor is expressed in terms of Ricci tensor
(and also the strength of $B_2$ is $H_{mnk} = H \ep_{mnk}$)
and thus it is possible that there is a choice of reparametrization that demonstrates also the two-loop renormalizability.}
i.e.
its form is preserved under the RG flow
with only the parameters $\ka,h,\bah$ and the overall scale $T$ in \rf{b1} changing ($p$ is not renormalized).
The IR fixed point corresponds to $\ka \to 0$ and thus to the marginal deformation of the $SU(2)$ WZW model
\rf{b12} which becomes a Weyl-invariant
sigma model when supplemented by an appropriate dilaton. The UV fixed point corresponds to $\ka \to 1$
when the model flows to the marginal deformation of the $SL(2,R)$ WZW background, which again represents a conformal sigma model.
Thus the RG flow connects the deformed $S^3$ and the euclidean $AdS_3$ or $H^3$ spaces just like
in the case of the simple symmetric 1-parameter deformed
coset model discussed in section 3 ($\k=0$ and $\k=\infty$ correspond to $\ka=0$ and $\ka=1$, see \rf{kaa1}).

To find the conformal sigma model representing the string solution
with the NS-NS background \rf{b1}, \rf{b2}, \rf{bb2} which may correspond to a deformation of the \ads supercoset
with a non-zero coefficient $\qq$ for the WZ term,
one would need to switch on also the RR background fields
(and determine the corresponding dilaton).

Finally, let us note that
one of the two parameters $h, \bah$ that controls the WZ coupling in \rf{b1}, \rf{b2}, \rf{bb2}
may be generated by a T-duality transformation. Since T-duality formally
preserves the integrability of the model (see, e.g., \ci{frt,fa,rtw}) the ``core'' integrable 3d sigma model
with two isometries may thus be characterized just by 3 parameters, that can be chosen, e.g.,
as the two parameters of Fateev model or $\ka$ and $p$ and the coefficient of the WZ term.
Indeed, performing the following TsT
transformation:\foot{This transformation is equivalent
to a non-trivial $O(2,2)$ duality transformation \ci{has}
depending on an $O(2)$ rotation matrix with angle $\a$ such that
$\g= -\tan \a$, provided one also rescales
the coordinates $\chi_i$ by $\cos \a$ and makes a constant shift of $B$ by $\g$.
}
T-duality $\chi_1\to \td \chi_1 $, shift of $\chi_2 \to \chi_2 + \gamma \td \chi_1$, and reverse T-duality
$\td \chi_1 \to \bar \chi_1$ gives a model of the same type as in \rf{b1} but with redefined functions
$D, \hat D, C, B$ containing one extra free parameter $\gamma$:
\be
&& D'= K^{-1} D\ , \ \ \ \ \hat D' = K^{-1} \hat D\ , \ \ \ C'= K^{-1} C\ , \ \ \ \
B' =K^{-1} \big[ B + \g (B^2 + \Delta ) \big] \ ,\ \ \ \la{b22} \\
&& K \equiv ( 1 + \g B)^2 + \g^2 \Delta \ , \ \ \ \ \ \ \ \ \ \
\Delta\equiv D \hat D - C^2 = 4 m^2 R^2 ( 1-z^2) ( 1 - \ka^2 z^2) Q^2 \ . \ \ \ \la{b23} \ee
The transformed functions have, in general, a different
dependence on $z$ as compared to the original one in \rf{b2}, but this
transformation is supposed to act on a special 3-parameter case to produce a
4-parameter one.
In particular, starting with the 1-parameter deformation case without $B$-term
($p=1, h =\bah=0$) and applying \rf{b22}, \rf{b23} one gets a special case of \rf{b1}, \rf{b2}, \rf{bb2} with non-zero $h=-\bah $.

\section{The $\varkappa=i$ action from the $AdS_2\times S^2$ supercoset \label{appC}}
\def\theequation{C.\arabic{equation}}
\setcounter{equation}{0}

In this Appendix we include some details of the construction of the deformed supercoset action in the
two $\varkappa\rightarrow i$ (or, equivalently, $\eta\rightarrow i$) limits. The details of the construction
depend quite strongly on the choice of $Z_4$ automorphism; it turns out that a convenient one is that of
\cite{gt1}, which identifies $\diag(\sigma_1,0)$ and $\diag(0,i\sigma_1)$ as the generators of the gauge group in the $AdS_2\times S^2$
supercoset.
With this choice, the coset representative takes the form
\be
\label{app_coset_rep}
&& g = g_B g_F \ , \qquad
g_B =
\begin{pmatrix}
g_A & 0_{2\times 2} \cr
0_{2\times 2} & g_S
\end{pmatrix} \ ,
\qquad
g_F =\exp F \ ,
\\ &&
g_A = e^{{1 \ov 2} it\sigma_3}
\begin{pmatrix}
\cosh a & i\sinh a \cr
-i \sinh a & \cosh a
\end{pmatrix} \ ,
\qquad
g_S = e^{{1\ov 2 } i\varphi \sigma_3}
\begin{pmatrix}
\cos b & \sin b \cr
-\sin b & \cos b
\end{pmatrix}\ .
\ee
Denoting by $Q_i$ and $S_i$ the $PSU(1,1|2)$ generators with charges $\pm i$, respectively,
\be
\begin{array}{cccc}
{
Q_0=
\left(\begin{smallmatrix}
0 & 0 & 1 & 0 \\
0 & 0 & 0 & 0 \\
i & 0 & 0 & 0 \\
0 & 0 & 0 & 0 \\
\end{smallmatrix} \right)_{\vphantom{\big|}}
}
&
Q_1=
{
\left(\begin{smallmatrix}
0 & 0 & 0 & i \\
0 & 0 & 0 & 0 \\
0 & 0 & 0 & 0 \\
1 & 0 & 0 & 0 \\
\end{smallmatrix}\right)
}
&
Q_2=
{
\left(
\begin{smallmatrix}
0 & 0 & 0 & 0 \\
0 & 0 & 1 & 0 \\
0 & -i & 0 & 0 \\
0 & 0 & 0 & 0 \\
\end{smallmatrix}
\right)
}
&
Q_3=
{
\left(
\begin{smallmatrix}
0 & 0 & 0 & 0 \\
0 & 0 & 0 & i \\
0 & 0 & 0 & 0 \\
0 & -1 & 0 & 0 \\
\end{smallmatrix}
\right)
}
\\
S_0=
{
\left(
\begin{smallmatrix}
0 & 0 & i & 0 \\
0 & 0 & 0 & 0 \\
1 & 0 & 0 & 0 \\
0 & 0 & 0 & 0 \\
\end{smallmatrix}
\right)
}
&
S_1=
{
\left(
\begin{smallmatrix}
0 & 0 & 0 & 1 \\
0 & 0 & 0 & 0 \\
0 & 0 & 0 & 0 \\
i & 0 & 0 & 0 \\
\end{smallmatrix}
\right)
}
&
S_2=
{
\left(
\begin{smallmatrix}
0 & 0 & 0 & 0 \\
0 & 0 & i & 0 \\
0 & -1 & 0 & 0 \\
0 & 0 & 0 & 0 \\
\end{smallmatrix}
\right)
}
&
S_3=
{
\left(
\begin{smallmatrix}
0 & 0 & 0 & 0 \\
0 & 0 & 0 & 1 \\
0 & 0 & 0 & 0 \\
0 & -i & 0 & 0 \\
\end{smallmatrix}
\right)
}
\end{array} \ ,
\ee
we choose the matrix $F$ defining the fermionic part of the coset representative as
\be\label{fparam}
F &=& (q_0 + \eta\, s_0 ) Q_0 + (q_1 - \eta\, s_1 ) Q_1 + (q_2 + \eta\, s_2 ) Q_2 + (q_3 - \eta\, s_3 ) Q_3
\cr
& +& (s_0 + \eta\, q_0 ) S_0 \, + (s_1 - \eta\, q_1 ) S_1 \, + (s_2 + \eta\, q_2 ) S_2 \, + (s_3 - \eta\, q_3 ) S_3 \ .
\label{Fpart}
\ee
Here $q_i$ and $s_i$ are real and $F$ obeys the reality condition outlined in Appendix C of \cite{gt1}. The
formal limit $\varkappa\rightarrow i$ or, equivalently $\eta\rightarrow i$, changes the
reality condition obeyed by $F$; such a change is hinted at \cite{hmh} by the expected
relation to the PR model for the $AdS_2\times S^2$ superstring and the fact that the
fundamental excitations change from magnons (in the GS theory) to solitons (in the PR theory).

With this choice of fields, the quadratic terms around the null geodesic $x^+=t+\varphi=\mu\tau$
are diagonal and manifestly exhibit the decoupling of $q_0, ~ q_3, ~s_0, ~s_3$. This is a consequence
of kappa symmetry. We will choose to fix it setting to zero the decoupled fields,
\be
q_0=0= q_3=0=s_0=0=s_3 \ .
\ee
This gauge, setting to zero the diagonal entries of the upper-right and lower-left $2\times 2$ blocks of
the purely fermionic terms in the coset representative, is the analog of the \adss lightcone gauge
around the null geodesic.

The construction of the action from eq.~\eqref{dee} is straightforward albeit tedious; for generic $\eta$
the resulting expression is quite lengthy but it simplifies in the $\eta\rightarrow i$ limit.

Taking the naive limit $\eta\rightarrow i$ (i.e. setting $\eta=i$ directly without additional rescalings) leads, after a change of coordinates
\be
a\rightarrow {\rm arctanh}(\tan a)\ ,
\qquad
b\rightarrow \arctan (\tanh b)\ ,
\label{flat_transverse}
\ee
to the flat space metric. The fermionic Lagrangian then describes four free massless fermions.

As discussed in section \ref{ads2 varkappa eq i}, the limit \eqref{lim} with $\k^2 = -1+\epsilon^2$ (i.e. $\eta=i(1-\epsilon+\dots )$) makes
contact with the PR theory for the GS string in $AdS_2\times S^2$~\cite{gt1}. In this limit, using the coset representative in eq.~\eqref{app_coset_rep}
and rescaling all fermions by the factor $(\eta-i)^{1/2}/(2\sqrt{\mu})$, the Lagrangian in eq.~\eqref{dee} becomes (we use $\eta_{ij}={\rm diag}(-1, 1)$ and
$\epsilon^{01}=1$):
\be
L &=& c_\eta\, ( L_B+L_F)\ ,
\\
L_B&=&2 \eta^{ij}\partial_i a \partial_ja +2\eta^{ij} \partial_i b\partial_j b
+ \frac{\mu^2}{4}(\cos4a-\cosh4b)\ ,
\\
L_F&=&-q_1\partial_+ q_1 - q_2\partial_+ q_2 - s_1\partial_- s_1 - s_2\partial_- s_2
\nonumber\\[2pt]
&-& 2\left(\frac{\sinh2 b}{\cos2 a + \cosh2 b} \partial_- a + \frac{\sin2 a}{\cos2 a + \cosh2 b}\partial_- b\right) s_1 s_2
\nonumber\\[3pt]
&+&2\left(\frac{\sinh2 b}{\cos2 a + \cosh2 b} \partial_+ a + \frac{\sin2 a}{\cos2 a + \cosh2 b}\partial_+ b\right)q_1 q_2
\nonumber\\[2pt]
&+&\mu \frac{\sin2 a \sinh2b}{\cos2 a + \cosh2 b}(q_1 s_2 - q_1 s_1)
\nonumber\\[2pt]
&-&\frac{\mu}{2}\,\frac{\cos4 a + 2 \cos2 a \cosh2 b + \cosh4 b}{\cos2 a + \cosh2 b}\ (q_1 s_1 + q_2 s_2) \ .
\ee
We notice that the connection-like terms on the second and third lines of $L_F$ are total derivatives, $\partial_\mp\arctan(\tan a \, \tanh b )$, and
thus they may be eliminated by opposite rotations in the planes $(q_1, q_2)$ and $(s_1, s_2)$:
\be
&& X = \arctan(\tan a \, \tanh b ) \ ,
\\
&& q_1 \rightarrow \cos X \; q_1 + \sin X \; q_2\ ,
\qquad
q_2\rightarrow -\sin X\; q_1 + \cos X \; q_2 \ ,
\cr
&& s_1 \rightarrow \cos X \; s_1 - \sin X \; s_2\ ,
\qquad
s_2\rightarrow \sin X\; s_1 + \cos X \; s_2 \ .
\nonumber
\ee
The resulting fermionic Lagrangian is
\be
L_F&=&-q_1\partial_+ q_1 - q_2\partial_+ q_2 - s_1\partial_- s_1 - s_2\partial_- s_2
\nonumber\\[2pt]
&+&\mu\Big[\cos2 a\cosh2 b\ (s_1 q_1 + s_2 q_2) + \sin2 a \sinh2b\ ( q_1 s_2 - q_2 s_1 ) \Big] \ .
\ee
The complete light-cone gauge-fixed deformed supercoset Lagrangian to quadratic order in fermions can then be mapped to the
Lagrangian of the PR model for the $AdS_2\times S^2$ superstring~\cite{gt1} by a double-Wick rotation and
identifying the fields as
\be
&&
a=\textstyle\frac{1}{2}\varphi
~,\quad
b=\textstyle\frac{1}{2}\phi
~,\qquad
q_1=\nu
~,\quad
q_2=\rho
~,\quad
s_1=\beta
~,\quad
s_2 = \gamma \ ,
\ee
and accounting for the difference in the definition of partial derivatives $\partial_\pm$ in \cite{gt1} (see footnote~\ref{def_gt_partial}).

We expect that a similar derivation, showing equivalence with the corresponding PR model, should be possible also for the
$AdS_3\times S^3$ deformed supercoset in the limit $\eta\rightarrow i$.



\newpage
\begin{thebibliography}{30}
\parskip=0.2 pt

\baselineskip 13pt

\bi{beir}
N.~Beisert, C.~Ahn, L.~F.~Alday, Z.~Bajnok, J.~M.~Drummond, L.~Freyhult, N.~Gromov and R.~A.~Janik {\it et al.},
``Review of AdS/CFT Integrability: An Overview,''
Lett.\ Math.\ Phys.\ {\bf 99}, 3 (2012)
[arXiv:1012.3982].

\bi{lm}
O.~Lunin and J.~M.~Maldacena,
``Deforming field theories with U(1) x U(1) global symmetry and their gravity duals,''
JHEP {\bf 0505}, 033 (2005)
[hep-th/0502086].

\bi{frt}
S.~A.~Frolov, R.~Roiban and A.~A.~Tseytlin,
``Gauge-string duality for superconformal deformations of N=4 super Yang-Mills theory,''
JHEP {\bf 0507}, 045 (2005)
[hep-th/0503192].

\bi{fa}
S.~Frolov,
``Lax pair for strings in Lunin-Maldacena background,''
JHEP {\bf 0505}, 069 (2005)
[hep-th/0503201].
L.~F.~Alday, G.~Arutyunov and S.~Frolov,
``Green-Schwarz strings in TsT-transformed backgrounds,''
JHEP {\bf 0606}, 018 (2006)
[hep-th/0512253].

\bi{rtw}
R.~Ricci, A.~A.~Tseytlin and M.~Wolf,
``On T-Duality and Integrability for Strings on AdS Backgrounds,''
JHEP {\bf 0712}, 082 (2007)
[arXiv:0711.0707].

\bibitem{bei}
N.~Beisert, R.~Ricci, A.~A.~Tseytlin and M.~Wolf,
``Dual Superconformal Symmetry from \adss Superstring Integrability,''
Phys.\ Rev.\ D {\bf 78}, 126004 (2008)
[arXiv:0807.3228].

\bibitem{dmv}
F.~Delduc, M.~Magro and B.~Vicedo,
``An integrable deformation of the $AdS_5 \times S^5$ superstring action,''
Phys.\ Rev.\ Lett.\ {\bf 112}, 051601 (2014)
[arXiv:1309.5850].

\bibitem{dmv1}
F.~Delduc, M.~Magro and B.~Vicedo,
``On classical $q$-deformations of integrable sigma-models,''
JHEP {\bf 1311}, 192 (2013)
[arXiv:1308.3581].

\bibitem{k1}
C.~Klimcik,
``Yang-Baxter sigma models and dS/AdS T duality,''
JHEP {\bf 0212}, 051 (2002)
[hep-th/0210095].

\bibitem{k2}
C.~Klimcik,
``On integrability of the Yang-Baxter sigma-model,''
J.\ Math.\ Phys.\ {\bf 50}, 043508 (2009)
[arXiv:0802.3518].

\bibitem{abf}
G.~Arutyunov, R.~Borsato and S.~Frolov,
``S-matrix for strings on $\eta$-deformed $AdS_5 \times S^5$,''
[arXiv:1312.3542].

\bibitem{km}
I.~Kawaguchi, T.~Matsumoto and K.~Yoshida,
``The classical origin of quantum affine algebra in squashed sigma models,''
JHEP {\bf 1204}, 115 (2012)
[arXiv:1201.3058].
I.~Kawaguchi and K.~Yoshida,
``A deformation of quantum affine algebra in squashed WZNW models,''
[arXiv:1311.4696].

\bi{jor1}
I.~Kawaguchi, T.~Matsumoto and K.~Yoshida,
``Jordanian deformations of the \adss superstring,''
[arXiv:1401.4855].
``A Jordanian deformation of AdS space in type IIB supergravity,''
[arXiv:1402.6147].

\bi{bk}
N.~Beisert and P.~Koroteev,
``Quantum Deformations of the One-Dimensional Hubbard Model,''
J.\ Phys.\ A {\bf 41}, 255204 (2008)
[arXiv:0802.0777].
\bi{beis}
N.~Beisert,
``The Classical Trigonometric r-Matrix for the Quantum-Deformed Hubbard Chain,''
J.\ Phys.\ A {\bf 44}, 265202 (2011)
[arXiv:1002.1097].

\bibitem{hhmphase}
B.~Hoare, T.~J.~Hollowood and J.~L.~Miramontes,
``q-Deformation of the $AdS_5 \times S^5$ Superstring S-matrix and its Relativistic Limit,''
JHEP {\bf 1203}, 015 (2012)
[arXiv:1112.4485].

\bi{hul}
C.~M.~Hull,
``Timelike T duality, de Sitter space, large N gauge theories and topological field theory,''
JHEP {\bf 9807}, 021 (1998)
[hep-th/9806146].

\bibitem{gt1}
M.~Grigoriev and A.~A.~Tseytlin,
``Pohlmeyer reduction of \adss superstring sigma model,''
Nucl.\ Phys.\ B {\bf 800}, 450 (2008)
[arXiv:0711.0155].
\bibitem{ms}
A.~Mikhailov and S.~Schafer-Nameki,
``Sine-Gordon-like action for the Superstring in \adss,''
JHEP {\bf 0805}, 075 (2008)
[arXiv:0711.0195].
\bibitem{gt2}
M.~Grigoriev and A.~A.~Tseytlin,
``On reduced models for superstrings on $AdS_n \times S^n$,''
Int.\ J.\ Mod.\ Phys.\ A {\bf 23}, 2107 (2008)
[arXiv:0806.2623].

\bibitem{ht1}
B.~Hoare and A.~A.~Tseytlin,
``Tree-level S-matrix of Pohlmeyer reduced form of \adss superstring theory,''
JHEP {\bf 1002}, 094 (2010)
[arXiv:0912.2958].

\bi{ht2}
B.~Hoare and A.~A.~Tseytlin,
``Towards the quantum S-matrix of the Pohlmeyer reduced version of \adss superstring theory,''
Nucl.\ Phys.\ B {\bf 851}, 161 (2011)
[arXiv:1104.2423].

\bibitem{bab}
A.~Babichenko, B.~Stefanski, Jr. and K.~Zarembo,
``Integrability and the AdS(3)/CFT(2) correspondence,''
JHEP {\bf 1003}, 058 (2010)
[arXiv:0912.1723].

\bibitem{beb}
N.~Berkovits, M.~Bershadsky, T.~Hauer, S.~Zhukov and B.~Zwiebach,
``Superstring theory on \adt as a coset supermanifold,''
Nucl.\ Phys.\ B {\bf 567}, 61 (2000)
[hep-th/9907200].

\bibitem{fat}
V.~A.~Fateev,
``The sigma model (dual) representation for a two-parameter family of integrable quantum field theories,''
Nucl.\ Phys.\ B {\bf 473}, 509 (1996).

\bibitem{luk}
S.~L.~Lukyanov,
``The integrable harmonic map problem versus Ricci flow,''
Nucl.\ Phys.\ B {\bf 865}, 308 (2012)
[arXiv:1205.3201].

\bibitem{foz}
V.~A.~Fateev, E.~Onofri and A.~B.~Zamolodchikov,
``The Sausage model (integrable deformations of O(3) sigma model),''
Nucl.\ Phys.\ B {\bf 406}, 521 (1993).

\bibitem{k3}
C.~Klimcik,
``Integrability of the bi-Yang-Baxter sigma-model,''
[arXiv:1402.2105].

\bibitem{che}
I.~V.~Cherednik,
``Relativistically Invariant Quasiclassical Limits of Integrable Two-dimensional Quantum Models,''
Theor.\ Math.\ Phys.\ {\bf 47}, 422 (1981)
[Teor.\ Mat.\ Fiz.\ {\bf 47}, 225 (1981)].

\bi{kaz}
A.~Cagnazzo and K.~Zarembo,
``B-field in AdS(3)/CFT(2) Correspondence and Integrability,''
JHEP {\bf 1211}, 133 (2012)
[Erratum-ibid.\ {\bf 1304}, 003 (2013)]
[arXiv:1209.4049].

\bi{mm}
J.~M.~Maldacena and L.~Maoz,
``Strings on pp waves and massive two-dimensional field theories,''
JHEP {\bf 0212}, 046 (2002)
[hep-th/0207284].

\bi{rt}
J.~G.~Russo and A.~A.~Tseytlin,
``A class of exact pp wave string models with interacting light cone gauge actions,''
JHEP {\bf 0209}, 035 (2002)
[hep-th/0208114].

\bibitem{bs}
I.~Bakas and J.~Sonnenschein,
``On Integrable models from pp wave string backgrounds,''
JHEP {\bf 0212}, 049 (2002)
[hep-th/0211257].

\bi{gh}
M.~T.~Grisaru, P.~S.~Howe, L.~Mezincescu, B.~Nilsson and P.~K.~Townsend,
``N=2 Superstrings in a Supergravity Background,''
Phys.\ Lett.\ B {\bf 162}, 116 (1985).

\bibitem{Zarembo:2010sg}
K.~Zarembo,
``Strings on Semi symmetric Superspaces,''
JHEP {\bf 1005}, 002 (2010)
[arXiv:1003.0465].

\bi{hu}
M.~Blau, J.~M.~Figueroa-O'Farrill, C.~Hull and G.~Papadopoulos,
``Penrose limits and maximal supersymmetry,''
Class.\ Quant.\ Grav.\ {\bf 19}, L87 (2002)
[hep-th/0201081].

\bibitem{bmn}
D.~E.~Berenstein, J.~M.~Maldacena and H.~S.~Nastase,
``Strings in flat space and pp waves from N=4 superYang-Mills,''
JHEP {\bf 0204} (2002) 013
[hep-th/0202021].

\bi{van} 
G.~Arutynov, M.~de Leeuw and S.~J.~van Tongeren,
``On the exact spectrum and mirror duality of the $(AdS_5 \times S^5)_\eta$ superstring,''
[arXiv:1403.6104].

\bi{AF}
G.~Arutyunov and S.~Frolov,
``On String S-matrix, Bound States and TBA,''
JHEP {\bf 0712}, 024 (2007)
[arXiv:0710.1568].

\bibitem{alt}
G.~Arutyunov, M.~de Leeuw and S.~J.~van Tongeren,
``The Quantum Deformed Mirror TBA I,''
JHEP {\bf 1210} (2012) 090
[arXiv:1208.3478];
``The Quantum Deformed Mirror TBA II,''
JHEP {\bf 1302} (2013) 012
[arXiv:1210.8185].

\bi{hmh}
B.~Hoare, T.~J.~Hollowood and J.~L.~Miramontes,
``A Relativistic Relative of the Magnon S-Matrix,''
JHEP {\bf 1111}, 048 (2011)
[arXiv:1107.0628].

\bi{hhmunit}
B.~Hoare, T.~J.~Hollowood and J.~L.~Miramontes,
``Restoring Unitarity in the q-Deformed World-Sheet S-Matrix,''
JHEP {\bf 1310}, 050 (2013)
[arXiv:1303.1447].

\bi{sft}
K.~Sfetsos and A.~A.~Tseytlin,
``Four-dimensional plane wave string solutions with coset CFT description,''
Nucl.\ Phys.\ B {\bf 427}, 245 (1994)
[hep-th/9404063].

\bi{hhmbs}
B.~Hoare, T.~J.~Hollowood and J.~L.~Miramontes,
``Bound States of the q-Deformed \adss Superstring S-matrix,''
JHEP {\bf 1210} (2012) 076
[arXiv:1206.0010].

\bi{htq}
B.~Hoare and A.~A.~Tseytlin,
``On string theory on $AdS_3 \times S^3 \times T^4$ with mixed 3-form flux: Tree-level S-matrix,''
Nucl.\ Phys.\ B {\bf 873}, 682 (2013)
[arXiv:1303.1037].

\bi{sq}
R.~Squellari,
``Yang-Baxter $\sigma$ model: Quantum aspects,''
[arXiv:1401.3197].

\bibitem{goi}
M.~Goykhman and E.~Ivanov,
``Worldsheet Supersymmetry of Pohlmeyer-Reduced $AdS_n \times S^n$ Superstrings,''
JHEP {\bf 1109}, 078 (2011)
[arXiv:1104.0706].

\bibitem{hom}
T.~J.~Hollowood and J.~L.~Miramontes,
``The $AdS_5 \times S_5$ Semi-Symmetric Space Sine-Gordon Theory,''
JHEP {\bf 1105}, 136 (2011)
[arXiv:1104.2429].
\bibitem{sh}
D.~M.~Schmidtt,
``Integrability vs Supersymmetry: Poisson Structures of The Pohlmeyer Reduction,''
JHEP {\bf 1111}, 067 (2011)
[arXiv:1106.4796].

\bibitem{sor}
D.~Sorokin, A.~Tseytlin, L.~Wulff and K.~Zarembo,
``Superstrings in $AdS_2 \times S^2 \times T^6$,''
J.\ Phys.\ A {\bf 44}, 275401 (2011)
[arXiv:1104.1793].

\bibitem{ros}
R.~Roiban and W.~Siegel,
``Superstrings on \adss supertwistor space,''
JHEP {\bf 0011}, 024 (2000)
[hep-th/0010104].

\bibitem{holm}
T.~J.~Hollowood and J.~L.~Miramontes,
``Symplectic Deformations of Integrable Field Theories and AdS/CFT,''
[arXiv:1403.1899].


\bi{tse}
 A.~A.~Tseytlin,
  ``On A 'Universal' class of WZW type conformal models,''
  Nucl.\ Phys.\ B {\bf 418}, 173 (1994)
  [hep-th/9311062].


\bibitem{sfet}
K.~Sfetsos,
``Integrable interpolations: From exact CFTs to non-Abelian T-duals,''
[arXiv:1312.4560].

\bibitem{hoh}
J.~H.~Horne and G.~T.~Horowitz,
``Exact black string solutions in three-dimensions,''
Nucl.\ Phys.\ B {\bf 368}, 444 (1992)
[hep-th/9108001].

\bi{has}
A.~Giveon and M.~Rocek,
``Generalized duality in curved string backgrounds,''
Nucl.\ Phys.\ B {\bf 380}, 128 (1992)
[hep-th/9112070].
S.~F.~Hassan and A.~Sen,
``Marginal deformations of WZNW and coset models from O(d,d) transformation,''
Nucl.\ Phys.\ B {\bf 405}, 143 (1993)
[hep-th/9210121].
M.~Henningson and C.~R.~Nappi,
``Duality, marginal perturbations and gauging,''
Phys.\ Rev.\ D {\bf 48}, 861 (1993)
[hep-th/9301005].
E.~Kiritsis,
``Exact duality symmetries in CFT and string theory,''
Nucl.\ Phys.\ B {\bf 405}, 109 (1993)
[hep-th/9302033].

\bi{hort}
G.~T.~Horowitz and A.~A.~Tseytlin,
``On exact solutions and singularities in string theory,''
Phys.\ Rev.\ D {\bf 50}, 5204 (1994)
[hep-th/9406067].

\bibitem{for}
S.~Forste,
``A Truly marginal deformation of SL(2, R) in a null direction,''
Phys.\ Lett.\ B {\bf 338}, 36 (1994)
[hep-th/9407198].

\end{thebibliography}
\end{document}